\documentclass[11pt,a4paper,fleqn]{article}
\usepackage[utf8]{inputenc}
\usepackage{authblk}
\usepackage{geometry}
\usepackage{xcolor}
\usepackage[round]{natbib}
\usepackage{graphicx}
\usepackage{pdflscape}  
\usepackage{appendix}
\usepackage{threeparttable}
\usepackage{booktabs}
\usepackage{float}
\usepackage{amssymb}
\usepackage{amsbsy}
\usepackage{bm}
\usepackage{amsmath}
\usepackage{amsthm}
\usepackage{graphicx}
\usepackage{mathtools}
\usepackage{rotating}
\usepackage{setspace}

\usepackage{color, colortbl}
\definecolor{ashgrey}{rgb}{0.7, 0.75, 0.71}
\definecolor{columbiablue}{rgb}{0.61, 0.87, 1.0}
\definecolor{coral}{rgb}{1.0, 0.5, 0.31}

\definecolor{colBVAR}{HTML}{bababa}
\definecolor{colBART}{HTML}{d7191c}
\definecolor{colmixBART}{HTML}{fdae61}
\definecolor{colerrorBART}{HTML}{abd9e9}
\definecolor{colfullBART}{HTML}{2c7bb6}

\definecolor{colcons}{HTML}{e31a1c}
\definecolor{colSV}{HTML}{a6cee3}
\definecolor{colhBART}{HTML}{1f78b4}

\usepackage{enumitem}
\newlist{steps}{enumerate}{1}
\setlist[steps,1]{label = Step \arabic*:}

\usepackage{adjustbox}
\usepackage{dcolumn}
\usepackage[justification=centering]{caption}

\newcolumntype{d}[1]{D..{#1}} % for alignment of numbers on decimal marker

\usepackage{caption}
\usepackage{subcaption}
\definecolor{nblue}{HTML}{000660}
\usepackage[colorlinks=true,urlcolor=nblue,linkcolor=nblue,citecolor=nblue]{hyperref}

\newcommand*{\myeqref}[2][Eq.~]{%
  \hyperref[{#2}]{#1(\ref*{#2})}%
}
\def\equationautorefname#1#2\null{%
  Eq.#1(#2\null)%
}
\newcommand{\argmin}{\arg\!\min}

\begin{document}
\title{\textbf{Forecasting US Inflation Using Bayesian Nonparametric Models}\thanks{
%\textit{Corresponding author}: Name. Affiliation. \textit{Address}: Address. \textit{Email}: \href{mailto:name@something.com}{name@something.com}. 
We are grateful to Michael Pfarrhofer and Niko Hauzenberger for useful comments and suggestions. The views expressed herein are solely those of the authors and do not necessarily reflect the views of the Federal Reserve Bank of Cleveland or the Federal Reserve System. Huber gratefully acknowledges financial support from the Austrian Science Fund (FWF, grant no. ZK 35).}}

\author[a]{Todd E. \textsc{Clark}}
\author[b]{Florian \textsc{Huber}}
\author[c]{Gary \textsc{Koop}}
\author[d]{Massimiliano \textsc{Marcellino}}
\affil[a]{\textit{Federal Reserve Bank of Cleveland}}
\affil[b]{\textit{University of Salzburg}}
\affil[c]{\textit{University of Strathclyde}}
\affil[d]{\textit{Bocconi University, IGIER and CEPR}}
\date{\today}

\maketitle\thispagestyle{empty}\normalsize\vspace*{-2em}\small\linespread{1.5}
\begin{center}
\begin{minipage}{0.8\textwidth}
\noindent\small The relationship between inflation and predictors such as unemployment is potentially nonlinear with a strength that varies over time, and prediction errors error may be subject to large, asymmetric shocks. Inspired by these concerns, we develop a model for inflation forecasting that is nonparametric both in the conditional mean and in the error using Gaussian and Dirichlet processes, respectively. We discuss how both these features may be important in producing accurate forecasts of inflation. In a forecasting exercise involving CPI inflation, we find that our approach has substantial benefits, both overall and in the left tail, with nonparametric modeling of the conditional mean being of particular importance.      

\end{minipage}
\end{center}

\normalsize\newpage
\section{Introduction}

%\textbf{update/recycle to GP-BNP}

As reviewed in studies including \citet{StockWatson2007inflation} and \citet{FaustWright2013inflation}, inflation forecasting is fraught with challenges.  Structural economic models and simple economic reasoning imply that inflation should be forecastable with a range of indicators, including measures of domestic and international economic activity, import prices or exchange rates, cost measures such as wage growth, and oil prices.  Other work has explored forecasting inflation with forward-looking financial indicators, such as bond yields.  Over some periods and in some studies, some of these variables have yielded some success in improving the accuracy of inflation forecasts.  However, by now a body of work has established that simple inflation forecasts from the unobserved components model of \citet{StockWatson2007inflation}  and the inflation gap model of \citet{FaustWright2013inflation} are very difficult for another specification to beat.  These specifications improve inflation forecasts by accounting for a time-varying trend in inflation.  But adding other information does little to help inflation forecasts.  For example, models motivated by the Phillips curve to include indicators of economic activity or marginal costs of production cannot consistently improve on these simple univariate benchmarks.\footnote{Although the Phillips curve does not appear to be broadly successful in out-of-sample forecasting of inflation, studies have documented some patterns in inflation dynamics consistent with the Phillips curve.  One such pattern, documented in \citet{StockWatson2010inflation}, is that inflation regularly falls around recessions.}

Although most of the literature alluded to has focused on parametric linear models of inflation, other work has examined the predictability of inflation using nonlinear or nonparametric models. For example, some work has suggested that the nonlinear effects of economic activity on inflation kick in as the economy becomes very strong, such that economic expansions fail to produce much of a rise in inflation until such times.  \citet{BabbDetmeister2017} provide a useful review of the literature on nonlinear Phillips curves.  A fast-growing literature evaluates the use of machine learning techniques for macroeconomic forecasting, with random forests (see \citet{breiman2001random} and, e.g., \citet{Masini2021machinelearning}, for a survey) performing particularly well, also during crisis times, in a variety of studies and for key variables such as GDP growth and inflation; see, e.g., \citet{Coulombe2020randomforest}, \citet{Coulombe2020machinelearning}, \citet{Coulombe2021COVID}, and \citet{medeiros2021forecasting}.  

While these papers adopt classical methods, Bayesian techniques are also available. \cite{jochmann2015modeling} proposes an infinite hidden Markov model and applies it to model US inflation dynamics. This model endogenously selects the number of regimes and, for US data, finds a secular decline in inflation volatility and around 7 distinct inflation regimes. In a recent paper, \cite{CHKMP2021} use multivariate Bayesian additive regression tree (BART) models to forecast several US macroeconomic aggregates (including inflation), with particular interest in forecasting tail risks. For inflation, BART models improve upon linear models in turbulent periods such as the COVID-19 pandemic.  

One shortcoming of BART (and in fact many nonparametric techniques such as Gaussian processes or kernel methods) is that the shocks are assumed to be Gaussian. If there is empirical evidence of non-Gaussian features such as heavy tails in the innovations, the flexibility of these nonparametric models could imply that the conditional mean captures this and the model would thus erroneously suggest nonlinear relationships between inflation dynamics and predictors of inflation.\footnote{This problem is closely related to the critique raised by \cite{sims2001evolving} about the model stipulated in \cite{cogley2001evolving}.} These considerations motivate the models we propose in this paper.

%This paper examines the ability of a particular class of nonparametric models to forecast inflation: Bayesian additive regression trees (BART; see \citet{chipman2010bart}).  These models provide a flexible and popular approach in many fields of statistics. \citet[HR]{huber2020inference} develop Bayesian methods that build BART into a VAR, leading to the Bayesian additive vector autoregressive tree model, and demonstrate that it forecasts well. \citet[][HKOPS]{huber2020nowcasting} develop Bayesian methods for the mixed-frequency version of this model, showing that it also forecasts well, particularly during the COVID-19 pandemic.  \citet{CHKMP2021} find that  BART (nonparametric) models improve the accuracy of forecasts of a set of US macroeconomic and financial indicators, with particular interest in forecasting tail risks.  In the inflation context, the flexibility of BART models for capturing nonlinearities may be helpful to forecasting inflation, by drawing out ear influences even when Phillips curve effects appear weak or uneven in linear models, or capturing complex dependencies on other indicators such as producer price inflation.

In this paper,  we consider models of inflation that combine nonparametric specifications of the conditional mean and nonparametric specifications of the innovation to inflation. We model the conditional mean using a flexible and analytically tractable Gaussian process (GP) regression. This specification is capable of capturing nonlinearities in the relationship between inflation and its predictors. To avoid overfitting, we introduce a subspace shrinkage prior \citep{subspace} that shrinks the GP regression toward a linear subspace in a data-driven manner. The resulting model can be interpreted as a convex combination between a GP regression and either a linear (estimated by OLS) or a factor model (when principal components rather than the original regressors are used to define the linear subspace).  To capture fat tails, possible asymmetries, and other non-Gaussian features that might determine inflation dynamics, we introduce a Dirichlet process mixture (DPM) model to estimate the unknown shock distribution. This mixture model allows us to capture unobserved heterogeneity in a very flexible manner and is thus capable of handling situations such as the pandemic. To assess which of these features improves inflation forecasts, we also consider variants that treat the conditional mean as linear or the error term as Gaussian. In addition, we  allow for error specifications with either constant variances or stochastic volatility.

After developing proper Markov chain Monte Carlo (MCMC) estimation algorithms, we use all of the models to forecast quarterly consumer price inflation (CPI) in the US, using various sets of predictors.  We evaluate out-of-sample forecasts over a long sample of 1980 through 2021, on the basis of the accuracy of point forecasts, density forecasts, and tail risk forecasts.

Our results confirm the benefits of our flexible, nonparametric   models.  Over the 1980 to 2021 period, our nonparametric models achieve some gains in the accuracy of point and density forecasts relative to the common benchmark of the univariate model of \citet{StockWatson2007inflation}.  These models achieve sizable gains during the volatile 2020-21 period of the pandemic.  The primary gains to flexible nonparametric modeling come from nonlinear modeling of the conditional mean, through Gaussian processes.  Although a large set of variables in some settings offers an advantage, it does not do so uniformly; most of the gains to nonparametric modeling can be achieved with a moderately sized set of variables.  In exercises that drill deeper into the properties of predictive distributions, our nonparametric models are also shown to yield gains in predicting left-tail risks to inflation.  They are more challenged in capturing time variation in the right tail, which may have to do with the prevalence of low and stable inflation for much of the sample.  Although the sample is small, the models seem to better capture the right tail of the predictive distribution of inflation during the pandemic period.  We also show that our proposed models yield predictive distributions that sometimes display asymmetry, related to the literature on inflation at risk; see, e.g., \citet{lopezloria}.

The paper proceeds as follows.  Section \ref{sec:model} presents our models and estimation algorithms.  Section \ref{sec:results} describes the data and  design of the forecasting exercise and presents results.  Section \ref{sec:coreinflation} summarizes results for a robustness check with an alternative measure of inflation that excludes food and energy.  Section \ref{sec:concl} concludes.

\section{A fully nonparametric model for forecasting inflation}\label{sec:model}
%\subsection{Overview}
Models used for macroeconomic forecasting involve assumptions about the functional and distributional form of the conditional mean and the form of the error process, respectively. In this paper we  use nonparametric forms for both of these, and it is in this sense that we refer to our model as fully nonparametric. For the conditional mean, we use a GP prior and for the error distribution a DPM model. In this section, we define our model, discuss its properties, and introduce MCMC methods that allow for computationally efficient Bayesian inference and prediction, with additional details presented in Appendix \ref{app:MCMC}. 

Our model assumes that inflation in time $t$, $y_{t}$, depends on a vector of $K$ appropriately lagged predictors, $\bm x_t$, in a possibly nonlinear way:
\begin{equation}
        y_{t}  = f(\bm x_t) + \varepsilon_{t}.
\label{GPregress}
\end{equation}
Here, $f: \mathbb{R}^K \to \mathbb{R}$ denotes an unknown and potentially nonlinear function. In the next sub-section, we focus on $f$. Then we present our  nonparametric treatment of $\varepsilon_t$.

\subsection{Nonparametric modeling of the conditional mean using GPs}
Learning the unknown function $f$ can be achieved through many different techniques such as Bayesian additive regression trees \citep{chipman2010bart}, B-splines \citep{subspace}, or (deep) neural networks \citep{nakamura2005inflation, coulombe2022neural}. 
In this paper, we propose approximating the unknown function $f$ using a GP regression. This approach places a GP prior on the function $f$. This implies a Gaussian prior on $\bm f = (f(\bm x_1),..,f(\bm x_T))'$ of the form:
\begin{equation}
\bm f \sim \mathcal{N}(\bm 0, \bm K),
\label{GPprior}
\end{equation}
with $\bm K$ being a $T \times T$ kernel matrix with typical element $k(\bm x_t, \bm x_\tau)$ for times $t$ and $\tau$.
        
GP priors are nonparametric in the sense that they do not assume a particular form for $f$; instead, they are interpreted as a prior over all functions that might fit the data. In essence, the $T$ elements in $\bm f$ are treated as unknown parameters. The likelihood defined by (\ref{GPregress}) is over-parameterized, but the use of prior information given in (\ref{GPprior}) can be used to overcome this concern. 
        
A textbook introduction to GPs is given in \cite{RasmussenWilliams}. A recent macroeconomic application is \cite{GPmacrouncertainty}, who also provide further intuition and explanation of GPs in an economic context. As compared to other approaches, GPs can be applied to data sets including many covariates without introducing additional parameters (and hence remaining relatively parsimonious). Another key advantage is that the computational burden is little affected by the number of covariates but depends largely on the number of observations. This makes GP regression well suited to quarterly macroeconomic data where $T$ is relatively small. 

Function estimation through GPs relies heavily on the particular choice of the kernel. Suitable kernels allow for capturing many different functional shapes and dynamics for the function $f$.  In principle, many choices for $\bm K$ are possible. In a time-series context, kernels can be developed for capturing low-frequency movements or abrupt breaks. In this sense, they can approximate the behavior of unobserved component models and successfully extract trend inflation. There is also a way of specifying them that leads to a specific form of a neural network; see \cite{DeepNeuralGP}. But the most common choice, which we also adopt in some cases, is the Gaussian kernel. A typical element of $\bm K$ under a Gaussian kernel is given by:
\begin{equation}
k(\bm x_t, \bm x_\tau) = \xi \times  \exp\left(- \frac{\phi}{2} ||\bm x_t - \bm x_\tau||^2 \right), \label{GaussianKernel}
\end{equation}
with $\xi, \phi \in \mathbb{R}^+$ denoting the hyperparameters of the kernel. This Gaussian kernel captures the idea that similar values for $\bm x_t$ and $\bm x_\tau$ should be associated with similar values for $f(\bm x_t)$ and $f(\bm x_\tau)$. The distance between two values of $\bm x_t$ and $\bm x_\tau$ is measured in squared exponential terms. The degree of smoothness of the function depends on $\phi$. Low values for this hyperparameter lead to a smooth function, whereas higher values allow for more high-frequency variation. Note that if  $\bm x_t = \bm x_\tau$, then $\text{Var}(f(\bm x_t))=\xi$. This allows us to see that $\xi$ controls the variance of the function $f$.  Since these hyperparameters are crucial for appropriately capturing inflation dynamics, we estimate them using a Bayesian approach.  This requires adequate priors. We found that values of $\xi$ and $\phi$ greater than $1$ led to overfitting, and thus, we use a Uniform prior between $0$ and $1$ to avoid values of $\xi, \phi >1$. This choice implies that, as long as these hyperparameters are not too large, we remain agnostic on the precise values of $\xi$ and $\phi$.

%Further details and the MCMC algorithm we use are given in the Appendix. 
    
We also introduce a second version of the GP prior, which involves the concept of subspace shrinkage; see \cite{subspace}. To motivate this addition, note that the GP prior using the Gaussian kernel reflects a belief in smoothness. However, we might  also be interested in a prior that reflects a belief in linearity. Linearity is a subspace of the nonlinear form of $f$, hence the terminology subspace shrinkage. This might be useful in and of itself (i.e., shrinking toward a more parsimonious model often improves forecasts), but subspace shrinkage methods can also be used as a model selection device (i.e., they can select the linear model if the data warrant this; see \citet{subspaceVAR}). 
    
\cite{subspace} show how one can shrink toward a pre-specified subspace such as the linear one in the context of a particular nonparametric model (in their case, a B-spline regression).\footnote{It is noteworthy that \cite{subspace}, in the working paper version, also use subspace shrinkage to force a GP regression toward a pre-specified parametric alternative.} Here we adapt these methods for our purposes. 

Let $\bm \Phi_0 = \bm X (\bm X' \bm X)^{-1}\bm X'$ denote the linear projection matrix of $\bm X = (\bm x'_1, \dots, \bm x'_T)'$. Subspace shrinkage involves modifying the prior variance, which in our case is the kernel, as follows: 
\begin{equation}
        \bm K_1 = \left( \bm K^{-1} + (\bm I - \bm \Phi_0)/\tau^2\right)^{-1}.
        \label{subspacekernel}
\end{equation}
The GP prior in (\ref{GPprior}) is replaced by
\begin{equation*}
        \bm f \sim \mathcal{N}(\bm 0, \bm K_1).
\end{equation*}
We stress that our GP subspace shrinkage prior is still a GP prior, but with a different choice of kernel. Hence, the same Bayesian MCMC methods such as those for the GP prior with Gaussian kernel described above can be used. The only addition is that we treat   $\tau^2$ as an unknown parameter. We assume that the prior on $\tau$ has a density proportional to:
\begin{equation*}
    p(\tau) \propto \frac{(\tau^2)^{d_1-1/2}}{(1+\tau^2)^{d_0+d_1}}\quad \text{ for } \tau \in (0, \infty).
\end{equation*}
This prior reduces to the half-Cauchy distribution if $d_0=d_1=1/2$, a choice we follow in this paper \citep{subspace}. 
    
The parameter $\tau^2$ plays an important role in the prior as it controls the weight on the linear part of the model. If $\tau^2$ is large, little weight is placed on the linear model. But as $\tau^2$ decreases, the linear part receives more weight. Formally,  Lemma 3.1 of \cite{subspace} shows that:
\begin{equation*}
        \mathbb{E}(\bm f| \omega, \bullet) =(1-\omega) \overline{\bm f} + \omega \bm \Phi_0 \bm Y,
\end{equation*}
with $\omega = 1/(1+\tau^2) \in [0, 1]$ and $\overline{\bm f}$ being the posterior mean of $\bm f$ for the GP process with Gaussian kernel.  This shows that the posterior fit of $\bm f$ under the subspace shrinkage prior can be interpreted as a convex combination between the fit of a GP regression with kernel $\bm K$ and the fit of a simple OLS regression.\footnote{If $K > T$, the inverse of $\bm X' \bm X$  does not exist. In this case, we replace $\bm \Phi_0 = \bm X (\bm X' \bm X)^{-1} \bm X'$  with $\bm \Phi_0 = \bm S (\bm S' \bm S)^{-1} \bm S'$, where $\bm S$ denotes a small number of principal components.} The same result also holds when interest centers on predictive distributions, thus implying that our approach can, in a data-driven way, assess whether inflation is better described by a linear model or whether nonlinearities in the conditional mean are necessary.
 
\subsection{Nonparametric modeling of the error distribution}
We now turn to the modeling of the error distribution. The GP specification on the conditional mean implies a great deal of flexibility in terms of capturing arbitrary functional relationships between the covariates in $\bm x_t$ and $y_t$. However, macroeconomic time series are also subject to, e.g., infrequent large shocks, conditional heteroskedasticity, and possible multi-modality of the shocks. To capture such features without strong a priori assumptions, we rely on DPMs. DPMs have long been used as a Bayesian nonparametric method for uncovering unknown distributions; see, e.g., \cite{EscobarWest}. We use an implementation as in \cite{FS_MW}.

The DPM is an infinite mixture of distributions. We assume the errors to be independent over time
\begin{equation}
\varepsilon_t \sim \sum_{j=1}^\infty w_{j} \mathcal{N}(\mu_j, \sigma_j^2),
\label{DPM}
\end{equation}
with $\sum_{j=1}^\infty w_{j} = 1$ and $w_j \ge 0 \ \forall \ j$. We use the standard stick-breaking representation of the weights, $w_j$, developed in \cite{Sethuraman1994}, to cast the mixture into a finite dimensional representation. The stick-breaking representation can be interpreted as a prior on the weights $w_j$ that depends on auxiliary quantities $\xi_j$ as follows:
\begin{equation}
    w_1 = \xi_1,\quad w_j = \xi_j \prod_{i=1}^{j-1} (1 - \xi_i),\quad \text{ for } j>1, \label{eq: stickbreaks}
\end{equation}
and each $\xi_i \sim \mathcal{B}(1, \alpha)$ is Beta distributed with $\alpha$ being a hyperparameter. This hyperparameter plays an important role in controlling the clustering behavior of the DPM, and we thus introduce a Gamma prior $\alpha \sim \mathcal{G}(2, 4)$, a choice suggested in \cite{EscobarWest}. 

For the component means $\mu_j$ we use Gaussian priors centered on zero with variance $\underline{v}_j=4$. Given the scale of our data, this introduces relatively little information and allows for sufficient flexibility for capturing outliers. On the component precision $\sigma_j^{-2}$, we use Gamma priors $\sigma_j^{-2} \sim \mathcal{G}(c_0, c_1)$ with $c_0=10$ and $c_1=5$. This choice ensures a proper prior with mean 2 and variance $0.4$, introducing sufficient information if one of the components includes only very few (or no) observations.

Intuitively speaking, this mixture specification soaks up any variation in $y_t$ not explained through the GP component in the conditional mean. Since our choice of the kernel implies that the GP captures smoothly varying trends in inflation (which are determined by $\bm x_t$), the mixture model will capture transitory and possibly large shocks to the trend that can also be asymmetric. 

In our empirical work, we present results using this form for the errors. But note that it assumes the errors to be independent over time, which is a potential drawback with time-series data. Especially if the time series feature volatility clustering, the DPM has problems capturing persistence in terms of the variance of a time series. Standard stochastic volatility (SV) models capture this through a persistent latent volatility component. To get the best of both worlds, we propose a version of the DPM that is capable of detecting volatility clusters. Several specifications have been proposed in the literature for combining DPMs with SV; see, for instance, \cite{JensenMaheu2010} and \cite{JensenMaheu2014}. But these specifications imply that standard Kalman filter-based algorithms are not applicable. To circumvent this issue, we consider a model,  which we label DPM-SV, that replaces (\ref{DPM}) with
\begin{equation}
\varepsilon_t \sim \sum_{j=1}^\infty w_{j} \mathcal{N}(\mu_j, \sigma_t^2) 
\end{equation}
and lets $\sigma_t^2$ follow a standard SV process, with $\log \sigma_t^2$ evolving according to an AR(1) process. We assume the same stick-breaking process for the weights. Thus, this model combines some of the flexibility of the DPM with the empirically desirable properties of the SV models.

\subsection{A brief sketch of the posterior simulator}
Estimation of all of these models can be carried out using  MCMC techniques. In Appendix \ref{app:MCMC} we provide additional details on the precise steps.  In principle, our MCMC algorithm samples the infinite mixture and the associated quantities conditional on $\bm f$. The precise algorithm is based on an auxiliary representation of the mixture model. Estimation of the DPM involves sampling the mixture weights, the auxiliary classification indicators, and the component means and variances. For all of these steps full conditional posterior distributions take a well-known form and can be simulated through Gibbs sampling steps. The conditional posterior of the hyperparameter $\alpha$ takes no well-known form and is simulated through a Metropolis-Hastings (MH) updating step. All of these steps are discussed in detail in Appendix \ref{app: dpm}.

Conditional on the DPM, we can sample from the posterior of $\bm f$, which takes a $T$-dimensional Gaussian form. The hyperparameter $\tau$ is simulated using the slice sampler.  The parameters determining the shape of the kernel $\phi$ and $\xi$ are obtained through an MH step. More details are provided in Appendix \ref{app: GP}.

We carry out posterior inference by repeating the algorithm 20,000 times and discarding the first 10,000 draws as burn-in. Based on full-sample results, the sampler mixes well, with inefficiency factors across all parameters and latent states of the model being well below 40.

\section{Forecasting US inflation using nonparametric models}\label{sec:results}

This section briefly describes the data, summarizes the models and design of the forecasting exercise, and presents results.

\subsection{Data}
We use quarterly data that range from  1959:Q1 to 2021:Q3 from the FRED-QD database developed in \citet{McCrackenNgFREDQD} and maintained by the Federal Reserve Bank of St.\ Louis.  As a measure of medium-term inflation expectations, we also use a 5-quarters-ahead inflation expectation from the Survey of Professional Forecasters, obtained from the website of the Federal Reserve Bank of Philadelphia.  We focus on forecasting CPI inflation, measured as  $(400/h) \ln(P_{t+h}/P_t)$ at horizon $h= 1,4$, with results on ex food and energy inflation provided in a robustness check.

We run all of the models, except for the unobserved components (UC) ones, which do not include any explanatory variables, using three different data sets: one that uses only lagged values of inflation as explanatory variables (labeled AR(1)), a moderately sized data set that includes $29$ variables,  and a large one involving 169 variables (which, as indicated below, we sometimes reduce to principal components to facilitate estimation).  As detailed in Appendix \ref{app:data}, the set with 29 variables includes an array of major macro indicators fitting into broad categories commonly considered as possible predictors of inflation.  These include various indicators of economic activity (e.g, growth in industrial production, growth in payroll employment, and the unemployment rate), growth in wages and unit labor costs, producer price inflation, and financial indicators (e.g., interest rates, stock returns, and growth in business loans).

\subsection{Model summary, acronyms and design of the forecasting exercise}
In our forecasting exercise, we consider a variety of different models and different data sets. We consider four different treatments of the conditional mean, four different treatments of the error distribution, and three different sets of variable.  In all implementations, the models include one lag of the explanatory variables.  In this sub-section we list the models and define the acronyms we use.

For the conditional mean, we have two versions of the Gaussian process with and without subspace shrinkage: GP-sub and GP, respectively. We also present results assuming $f$ is linear and use the acronym Linear for this. This model is estimated as a nested alternative of GP-sub with $\omega = 1$. We also produce results for UC models that involve only a time-varying intercept that follows a random walk.  This model (with a stochastic volatility specification; see below) serves as our main benchmark given its strong performance when forecasting quarterly inflation (see, e.g., \citet{StockWatson2007inflation} and \citet{StockWatson2010inflation}). 

For the error distribution we assess whether allowing for departures from Gaussianity and homoskedasticity in a flexible way pays off. To do so, we consider conventional Normal errors with constant variance (labeled Homosk.) and also with  SV. The former serves as a natural benchmark to assess the predictive gains from allowing for heteroskedasticity, whereas the latter has been shown to be helpful in improving the accuracy of both point and density forecasts  \citep{clark2011real}. These traditional assumptions on the shocks are contrasted with our comparable nonparametric versions: the DPM and DPM-SV defined in the preceding sub-section. 

This leads to 16 different models that combine a choice for the conditional mean with one for the error distribution. We estimate and forecast with all of these. We run all of the models, except for the UC ones, which do not include any explanatory variables, using the three different data sets described above. These data sets are also used to form the subspace toward which we shrink in our GP-sub models. It is worth stressing that for the large GP regressions, the regressors enter the model in an unrestricted manner. But for the linear regression model, since OLS estimation using all $175$ regressors of the large data set is infeasible, we follow \cite{stock2002macroeconomic} and use a linear model with the first $6$ principal components being explanatory variables. As described in Footnote 4, this also forms the subspace toward which we shrink the large-scale GP regressions. 

The design of our forecasting exercise is recursive using an expanding window of data. We use the period from 1980:Q1 to 2021:Q3 as our forecast evaluation period and produce forecasts for horizons $h=1$ and $h=4$. These forecasts are produced by lagging the elements in $\bm x_t$ appropriately (i.e., we compute direct forecasts).

As measures of predictive accuracy we focus on the mean squared forecast error (MSE) and the log predictive likelihood (i.e., log score, denoted LPL). To assess tail forecasting accuracy, we will focus on the quantile score (QS), associated with the tick loss function as in studies such as \cite{giacomini2005qscore}:
\begin{equation*}
\mbox{QS}_{p, t+h} = (y_{t+h} - Q_{p, t+h})(p - \mathbf{1}_{\{y_{t+h} \le
Q_{p,t+h}\}}),
\end{equation*}
where $Q_{p,t+h}$ is the forecast of quantile $p$, and the indicator function $\mathbf{1}_{\{y_{t+h} \le Q_{p,t+h}\}}$ has a value of 1 if the outcome is at or below the forecast quantile and 0 otherwise.

\subsection{Results}

%xxx need to take another look at where to put this point (wasn't so clear that it fit clearly where it was, IMO):   DPM, being independent over time, sometimes does not manage to capture clusters of large volatility. 

\autoref{tab: fullres} reports MSE and average LPL (in parentheses) results for forecast horizons of $1$ and $4$ quarters.  For the benchmark UC-SV specification, we report the MSE and LPL levels, whereas for all other models, we report ratios of MSEs relative to the benchmark (ratios less than 1 mean a model is more accurate than the UC-SV model) and differences in average LPL relative to the benchmark (positive entries mean a model is more accurate than the benchmark).  The table has four horizontal panels for, respectively, UC, univariate, moderately sized, and large models. For all models we report results for linear, GP, and GP-sub specifications (except UC) for the conditional mean, and homoskedastic, DPM, SV, and DPM-SV for the error process. Finally, the table has two vertical panels:  the left one for a long evaluation period ranging from 1980:Q1 until 2021:Q3, the right one for the short COVID pandemic period, ranging from 2020:Q1 to 2021:Q3. Naturally, the short length of the COVID period should be taken into proper consideration when assessing the results, but it is of course of major interest, since most standard econometric models had severe problems in tracking and forecasting economic variables during this period due to their wide swings. In particular, inflation first dropped significantly, almost to deflation, and then rebounded so strongly that it reached levels last seen in the early 1980s.  In contrast, the low and stable inflation that prevailed from the early 1990s until the Great Recession may create challenges in beating the forecast accuracy of the simple UC-SV benchmark forecast.

Starting with the UC models, the main finding is that the pandemic led to large increases in MSE for $h=1$ but not for $h=4$, while there were massive decreases in LPL for both horizons.  Among the UC specifications, the homoskedastic specification is the least accurate.  Focusing on the other UC specifications, for $h=1$, there are only small differences in MSE from the various error specifications over the long evaluation period, with some gains for DPM and DPM-SV (about 5-7 percent) during the pandemic period, which, however, turn into large losses (up to 54 percent) for $h=4$. For LPL, there are also small differences in general, but, during the pandemic DPM and DPM-SV present some small gains, indicating that these more complex specifications produced slightly more accurate density forecasts. Overall, it seems difficult to consistently beat the benchmark UC-SV model with more complex specifications of the error term.

\begin{table}[hbt!]
\centering
\scalebox{0.75}{
\begin{tabular}{llllllllll}
   \hline
     &\multicolumn{4}{c}{Full hold-out period} & & \multicolumn{4}{c}{Only pandemic observations}\\
    &\multicolumn{4}{c}{(1980:Q1-2021:Q3)} & & \multicolumn{4}{c}{(2020:Q1 to 2021:Q3)}\\\midrule
 & Homosk. & DPM & SV & DPM-SV &  & Homosk. & DPM & SV & DPM-SV \\ 
    \midrule
     \multicolumn{10}{c}{Unobserved components models}\\ %\midrule
  $h=1$ & 1.144 & 0.952 & 0.679 & 1.002 &  & 1.161 & 0.949 & 1.086 & 0.929 \\ 
   & (-0.196) & (0.011) & (-1.063) & (0.002) &  & (0.489) & (0.355) & (-2.252) & (0.456) \\ 
  $h=4$ & 1.072 & 0.96 & 0.595 & 0.997 &  & 1.58 & 1.366 & 0.601 & 1.541 \\ 
   & (-0.357) & (-0.237) & (-0.867) & (0.006) &  & (0.337) & (0.406) & (-1.689) & (0.21) \\ 
    \midrule
  \multicolumn{10}{c}{AR(1) models}\\ 
    $h=1$ &  &  &  &  &  &  &  &  &  \\ 
  Linear & 1.32 & 1.329 & 1.233 & 1.198 &  & 0.942 & 0.946 & 0.945 & 0.944 \\ 
   & (-0.362) & (-0.376) & (-0.085) & (-0.074) &  & (0.587) & (0.436) & (0.597) & (0.602) \\ 
  GP & 1.005 & 1.039 & 1.005 & 1.009 &  & 0.776 & 0.808 & 0.811 & 0.805 \\ 
   & (0.057) & (0.046) & (0.262) & (0.267) &  & (0.924) & (0.659) & (0.789) & (0.775) \\ 
  GP-sub & 1.372 & 1.429 & 1.307 & 1.285 &  & 0.906 & 0.89 & 0.914 & 0.913 \\ 
   & (-0.422) & (-0.486) & (-0.197) & (-0.143) &  & (0.66) & (0.641) & (0.598) & (0.64) \\ 
  $h=4$ &  &  &  &  &  &  &  &  &  \\ 
  Linear & 1.657 & 1.428 & 1.248 & 1.145 &  & 1.961 & 1.836 & 1.523 & 1.513 \\ 
   & (-0.739) & (-0.614) & (-0.502) & (-0.336) &  & (-0.359) & (-0.494) & (-1.427) & (-0.69) \\ 
  GP & 1.019 & 1.026 & 0.955 & 0.97 &  & 0.852 & 0.869 & 0.869 & 0.879 \\ 
   & (-0.044) & (-0.025) & (0.107) & (0.135) &  & (0.919) & (0.998) & (0.874) & (0.941) \\ 
  GP-sub & 1.552 & 1.505 & 1.398 & 1.35 &  & 1.867 & 1.779 & 1.558 & 1.518 \\ 
   & (-0.607) & (-0.635) & (-0.641) & (-0.481) &  & (-0.237) & (-0.366) & (-1.38) & (-0.826) \\ 
       \midrule
  \multicolumn{10}{c}{Moderately sized models}\\ 
  $h=1$ &  &  &  &  &  &  &  &  &  \\ 
  Linear & 0.975 & 0.947 & 1.067 & 1.056 &  & 0.844 & 0.861 & 0.849 & 0.854 \\ 
   & (0.11) & (0.142) & (0.095) & (0.118) &  & (0.754) & (0.628) & (0.682) & (0.613) \\ 
  GP & 0.941 & 0.953 & 0.938 & 0.937 &  & 0.868 & 0.934 & 0.899 & 0.924 \\ 
   & (0.195) & (0.221) & (0.333) & (0.33) &  & (0.883) & (0.846) & (0.739) & (0.877) \\ 
  GP-sub & 0.925 & 0.943 & 0.996 & 0.95 &  & 0.879 & 0.925 & 0.89 & 0.908 \\ 
   & (0.189) & (0.246) & (0.065) & (0.321) &  & (0.796) & (0.839) & (0.642) & (0.728) \\ 
  $h=4$ &  &  &  &  &  &  &  &  &  \\ 
  Linear & 1.196 & 1.285 & 0.935 & 0.996 &  & 1.308 & 1.333 & 1.105 & 1.079 \\ 
   & (-0.24) & (-0.317) & (0.029) & (-0.012) &  & (0.371) & (0.43) & (-0.233) & (-0.052) \\ 
  GP & 0.856 & 0.871 & 0.873 & 0.848 &  & 0.529 & 0.536 & 0.669 & 0.573 \\ 
   & (0.171) & (0.126) & (0.255) & (0.249) &  & (1.193) & (1.174) & (1.186) & (1.292) \\ 
  GP-sub & 0.952 & 0.999 & 0.927 & 0.959 &  & 0.737 & 0.784 & 0.732 & 0.885 \\ 
   & (0.111) & (0.059) & (0.177) & (0.162) &  & (1.135) & (1.061) & (1.031) & (1.057) \\ 
          \midrule
  \multicolumn{10}{c}{Large-scale models}\\ 
  $h=1$ &  &  &  &  &  &  &  &  &  \\ 
  Linear & 0.921 & 0.984 & 0.933 & 0.941 &  & 0.805 & 0.82 & 0.829 & 0.839 \\ 
   & (0.117) & (0.082) & (0.218) & (0.224) &  & (0.886) & (0.83) & (0.857) & (0.852) \\ 
  GP & 0.981 & 1.062 & 0.972 & 0.984 &  & 0.638 & 0.792 & 0.724 & 0.753 \\ 
   & (0.126) & (0.117) & (0.179) & (0.291) &  & (1.217) & (1.048) & (1.141) & (1.131) \\ 
  GP-sub & 0.928 & 1.012 & 0.933 & 0.943 &  & 0.753 & 0.783 & 0.794 & 0.809 \\ 
   & (0.156) & (0.118) & (0.332) & (0.336) &  & (1.052) & (1.008) & (0.985) & (0.963) \\ 
  $h=4$ &  &  &  &  &  &  &  &  &  \\ 
  Linear & 0.883 & 1.058 & 0.888 & 0.922 &  & 0.779 & 0.874 & 0.769 & 0.749 \\ 
   & (0.004) & (0.031) & (0.029) & (0.02) &  & (0.948) & (1.071) & (0.975) & (1.031) \\ 
  GP & 1.046 & 1.128 & 0.947 & 0.959 &  & 0.684 & 0.724 & 0.624 & 0.622 \\ 
   & (0.013) & (-0.035) & (0.254) & (0.162) &  & (1.038) & (1.036) & (1.302) & (1.244) \\ 
  GP-sub & 0.927 & 1.085 & 0.843 & 0.875 &  & 0.747 & 0.731 & 0.717 & 0.726 \\ 
   & (0.07) & (-0.017) & (0.247) & (0.218) &  & (1.056) & (1.068) & (1.299) & (1.214) \\ 
   \hline
\end{tabular}
}
\caption{MSE and Average LPL Results.  LPL results given in parentheses. Results are relative (MSE ratios or LPL differences) to the UC-SV benchmark.}\label{tab: fullres}
\end{table}

% not clear we can really make this attribution:  mostly due to comparably good performance of UC-SV in this period

Moving to the AR(1) models, over the full evaluation sample the linear AR(1) specification is worse than UC-SV in terms of MSE and LPL for both $h=1$ and $h=4$, in line with previous results in the literature. Interestingly, GP is overall comparable with UC-SV in terms of MSE and LPL, with SV and DPM-SV being slightly more accurate than UC-SV for $h=4$. GP-sub is, not surprisingly, more similar to linear AR(1). During the COVID period, the linear AR(1) is slightly to modestly better than UC-SV for $h=1$ but much worse for $h=4$, and the same happens for GP-sub. Yet, GP is much better than UC-SV for both $h=1$ and $h=4$ and for both MSE and LPL, with minor differences across error types. Overall, GP seems a better univariate model than UC-SV, since it performs similarly in normal times, in particular when complemented with SV or DPM-SV, and much better in problematic periods. 

Next, we consider the role of the variables included in the moderately sized multivariate model. Over the full evaluation sample, at the one-step-ahead horizon the linear specification (with alternative error specifications) yields only small to modest differences in MSE (plus or minus 5 percent depending on the error term) with respect to UC-SV.  For the same sample and the four-steps-ahead horizon, the linear model is much less accurate, beaten by the UC-SV benchmark in most cases.  Over the pandemic sample, at the one-step-ahead horizon the linear model achieves larger gains in forecast accuracy (across all error specifications), of roughly 15 percent in MSE and even larger in LPL.  At the four-steps-ahead horizon, the linear model continues to be dominated by the UC-SV baseline.  Comparing these linear model results with those for the linear AR(1) specification indicates that, under linearity, adding variables typically improves MSE and LPL accuracy, but not enough that the resulting model consistently improves on the UC-SV baseline, particularly at the longer forecast horizon.

%For $h=4$, there are instead losses in MSE of up to 33\%, also during the pandemic, and losses also for LPL in the full period, and gains that either shrink or disappear during the pandemic. When comparing with the univariate AR(1) model, there are instead gains in MSE and LPL for all horizons, error specifications, and evaluation periods. Hence, adding variables is helpful for predicting inflation within the class of linear models, but the resulting multivariate linear model is not generally better than UC-SV.

With the moderately sized set of variables, the GP and GP-sub specifications consistently improve on the forecast accuracy of the UC-SV benchmark.  In the full sample, at both horizons the MSEs of the moderately sized GP and GP-sub models are slightly to modestly lower than those for linear regression and UC-SV, with gains with respect to the UC-SV baseline of about 5-8 percent for $h=1$ and as much as 15 percent for $h=4$, and small differences across error specifications.  The moderately sized GP and GP-sub models also offer small to modest gains in LPL.  In the pandemic sample, the MSE and LPL gains of the moderately sized GP and GP-sub models are larger.  For $h=4$, by the MSE metric, GP is much better than not only UC-SV but also the linear moderately sized model and the AR(1) GP specification. GP's better performance at $h=4$ with respect to both linear and UC-SV could be due to the increased importance of nonlinearity, in the sense that if the relationship between inflation and the explanatory variables is nonlinear at $h=1$, it will, in general, be even more nonlinear at $h=4$. Overall, the moderately sized multivariate GP specification seems so far to be the preferred model, beating the benchmark for both MSE and LPL across both horizons and evaluation periods, with little difference across error specifications.
  
  %and univariate GP, and GP is better than GP-sub, with gains with respect to UC-SV of about 15\%, as well as with gains in LPL. For both measures, gains increase during the pandemic, reaching close to 50\% in terms of MSE with respect to UC-SV.

%, particularly for $h=4$, with MSE gains nearing 50\%.

%During the pandemic, the MSE is instead slightly higher than for linear but still better than UC-SV. In addition, there are LPL gains with respect to UC-SV, larger than those from the linear model and larger during the pandemic. Yet, during the pandemic the small GP and GP-sub models are generally not better than the corresponding univariate models.  

Finally, we consider the larger multivariate models and, in the interest of brevity, we focus on the large-scale GP models as compared to the moderately sized GP specifications.\footnote{The larger information set clearly helps the linear regression model when predicting inflation, in particular during the pandemic, with few differences across error specifications.}  With the nonparametric specifications, it is mostly --- but not uniformly --- the case that models using the moderately sized set of variables forecast as well as or better than models with the large-scale data set.  More specifically, over the long evaluation sample, at both forecast horizons, using the large set of variables offers no gains in MSE or LPL (in fact, for $h=4$, using the large set sometimes reduces accuracy) as compared to using the moderately sized set of variables --- i.e., nonlinearity seems to make the larger information set redundant.  But in the pandemic period, results are more mixed, depending on the forecast horizon.  For $h=1$, the large-scale GP model yields lower MSE and higher LPL as compared to the moderately sized GP model; in this case, large GP-homoskedastic is best, with MSE gains of about 36 percent with respect to UC-SV.  But for $h=4$, the reverse is true.  For both horizons, the ranking of GP and GP-sub, and of the various error specifications, is not clear-cut, but in general the differences are small.

Overall, we conclude from the point and density forecast evaluation that our nonparametric models can offer gains in forecast accuracy over a long sample but particularly during the volatile period of the pandemic.  With no clear-cut ranking for the different specifications, we believe, for the error term of the inflation model, that the primary gains to flexible nonparametric modeling come from nonlinear modeling of the conditional mean, through GPs (without a clear advantage or disadvantage to applying subspace shrinkage).  Although a large set of variables offers an advantage in some settings, it does not do so uniformly; most of the gains to nonparametric modeling can be achieved with a moderately sized set of variables.

%Overall, Table 1 suggests that during the longer evaluation period, for h=1 the large linear regression is best in terms of MSE and the large GP-sub DPM-SV best in terms of LPL, but various models are close; for $h=4$, the small GP DPM-SV is best for both MSE and LPL. The best model during the pandemic is instead the large GP-homoskedastic for $h=1$, and for $h=4$, small GP-homoskedastic is best by MSE and small GP-DPM-SV is best by  LPL. More generally, a larger model seems preferred at $h=1$ and a smaller one for $h=4$, likely due to difficulties in modeling the medium-term evolution of all the variables in larger systems; a nonlinear specification of the conditional mean often helps, in particular during the pandemic and similar problematic periods, with little differences between GP and GP-sub; no clear-cut ranking instead emerges in terms of error specifications.

\subsection{Forecast performance over time}
\begin{figure}[t!]
    \centering
    \begin{minipage}[c]{1\linewidth}
       \textbf{(a) 1-quarter-ahead}
    \end{minipage}
    \begin{minipage}[c]{1\linewidth}
           \includegraphics[scale=0.71]{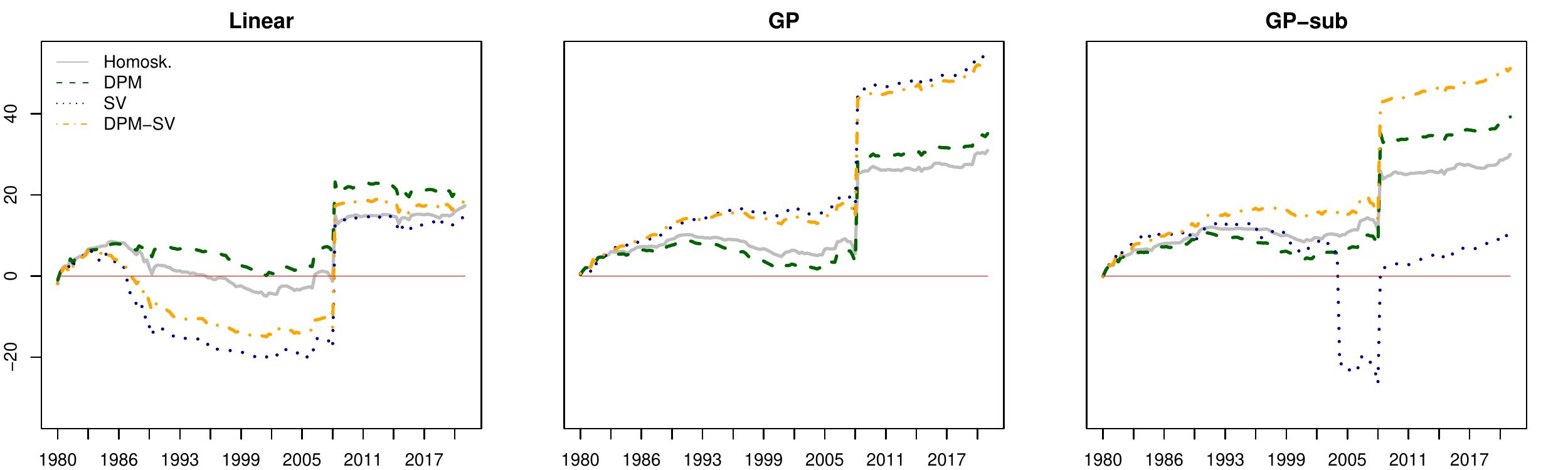}
    \end{minipage}
        \begin{minipage}[c]{1\linewidth}
        \vspace{0.2cm}
       \textbf{(a) 4-quarters-ahead}
    \end{minipage}
    \begin{minipage}[c]{1\linewidth}
           \includegraphics[scale=0.71]{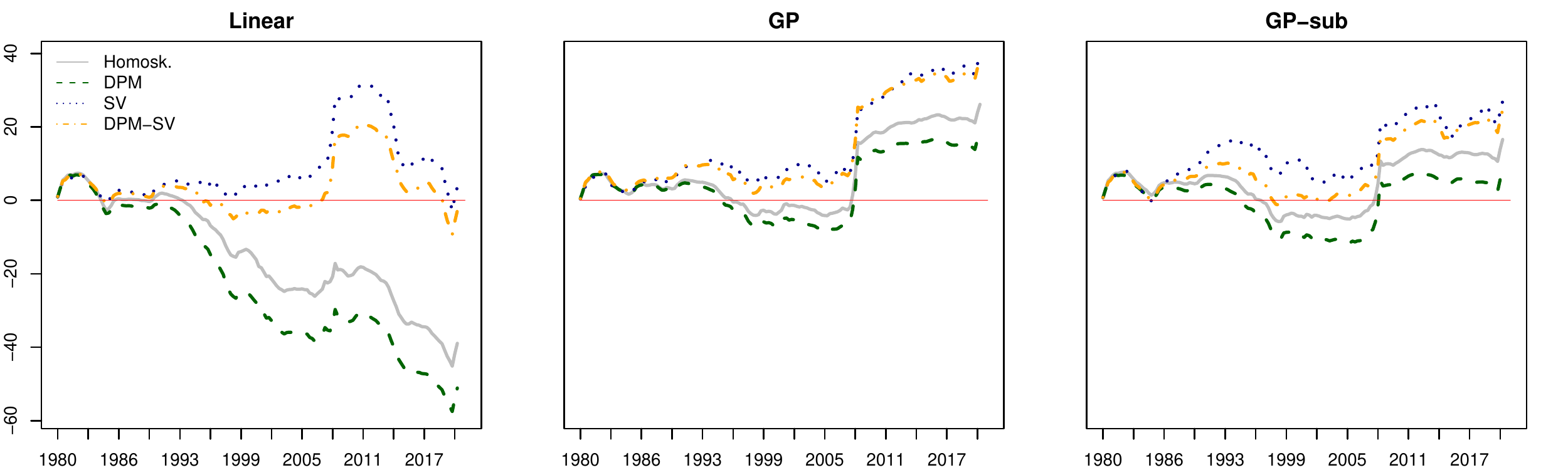}
    \end{minipage}
    \caption{Cumulative log predictive likelihoods against the UC-SV model: Moderately sized models}
    \label{fig:cumLPL_small}
\end{figure}

\begin{figure}[t]
    \centering
    \begin{minipage}[c]{1\linewidth}
       \textbf{(a) 1-quarter-ahead}
    \end{minipage}
    \begin{minipage}[c]{1\linewidth}
           \includegraphics[scale=0.71]{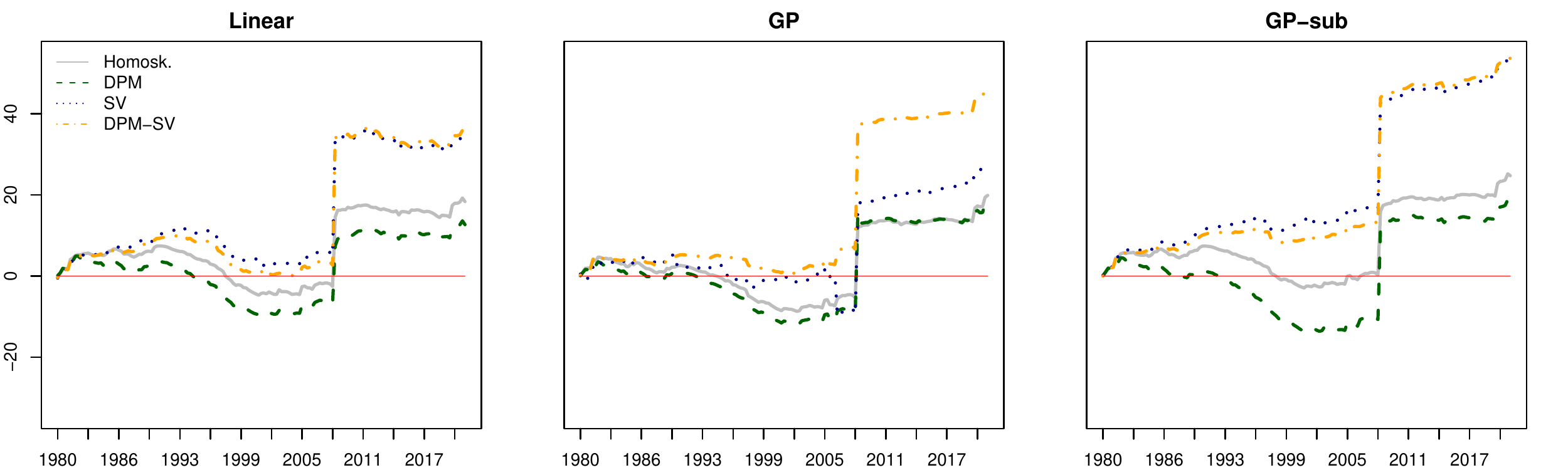}
    \end{minipage}
        \begin{minipage}[c]{1\linewidth}
        \vspace{0.2cm}
       \textbf{(a) 4-quarters-ahead}
    \end{minipage}
    \begin{minipage}[c]{1\linewidth}
           \includegraphics[scale=0.71]{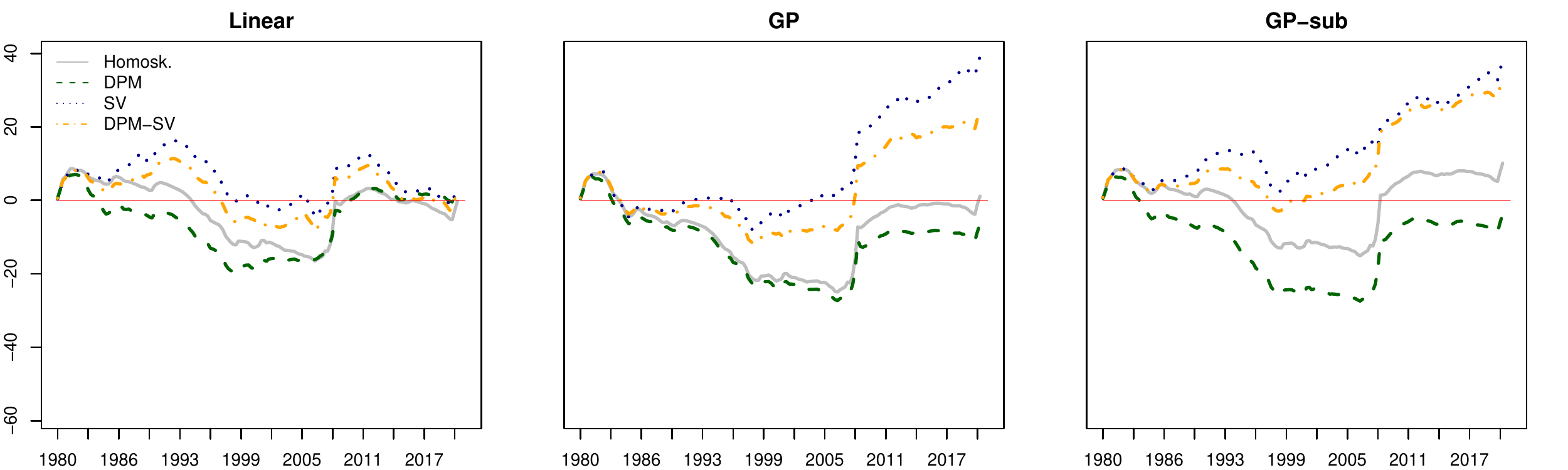}
    \end{minipage}
    \caption{Cumulative log predictive likelihoods against the UC-SV model: Large-scale models}
    \label{fig:cumLPL_large}
\end{figure}

We now discuss additional empirical results that may shed light on some of the patterns emerging from \autoref{tab: fullres}, and clarify additional aspects such as the sources of differences in LPLs and the stability of the relative model performance over time.  Figures \ref{fig:cumLPL_small} and \ref{fig:cumLPL_large} report cumulative differences in LPLs with respect to the UC-SV specification for, respectively, moderately sized and large models.  In each figure, the top and bottom rows provide results for $h=1$ and $h=4$, respectively.  Columns 1 to 3 report results for linear, GP, and GP-sub specifications, respectively.

%xxx to explain something I took out below:  it is not clear whether we refer to changes in scores with the crisis or whether we are referring to levels.  The changes in LPLS over time seem kind of comparable across models, so I made some cuts.  Also downplaying linear models.

Figures \ref{fig:cumLPL_small} and \ref{fig:cumLPL_large} show that the relative performance with respect to UC-SV improves around the financial crisis, more sharply for $h=1$ than for $h=4$.  Following the crisis, for the GP and GP-sub models applied to both the moderately sized and the large variable sets, the SV and DPM-SV error specifications are consistently first or second best in model fit, except in the case of the GP-sub specification at the one-step-ahead horizon.  By this measure, including stochastic volatility in the model's error specification improves model fit, particularly following the financial crisis.  Once again, in the GP class, the large-scale set of variables offers no clear advantages over the moderately sized data set.

%more so for GP and GP-sub than for linear, and for DPM-SV than DPM. The improvement remains also for $h=4$, though it shrinks. Figure \ref{fig:cumLPL_large} confirms that adding variables is useful for the linear model, in particular when combined with SV or DPM-SV. It can instead be useless or even detrimental for the GP models, for which DPM-SV or SV remain best. The improvement in performance with the financial crisis is evident also for the larger models, again more so for $h=1$ than for $h=4$.

%\subsection{Why does the Gaussian process model performs so well?}

\subsection{Tail forecasting performance}

\begin{figure}[t!]
\begin{minipage}[c]{\linewidth}
   \textbf{(a) Gaussian process}
\end{minipage}
   \begin{minipage}[c]{\linewidth}
   \includegraphics[scale=0.58]{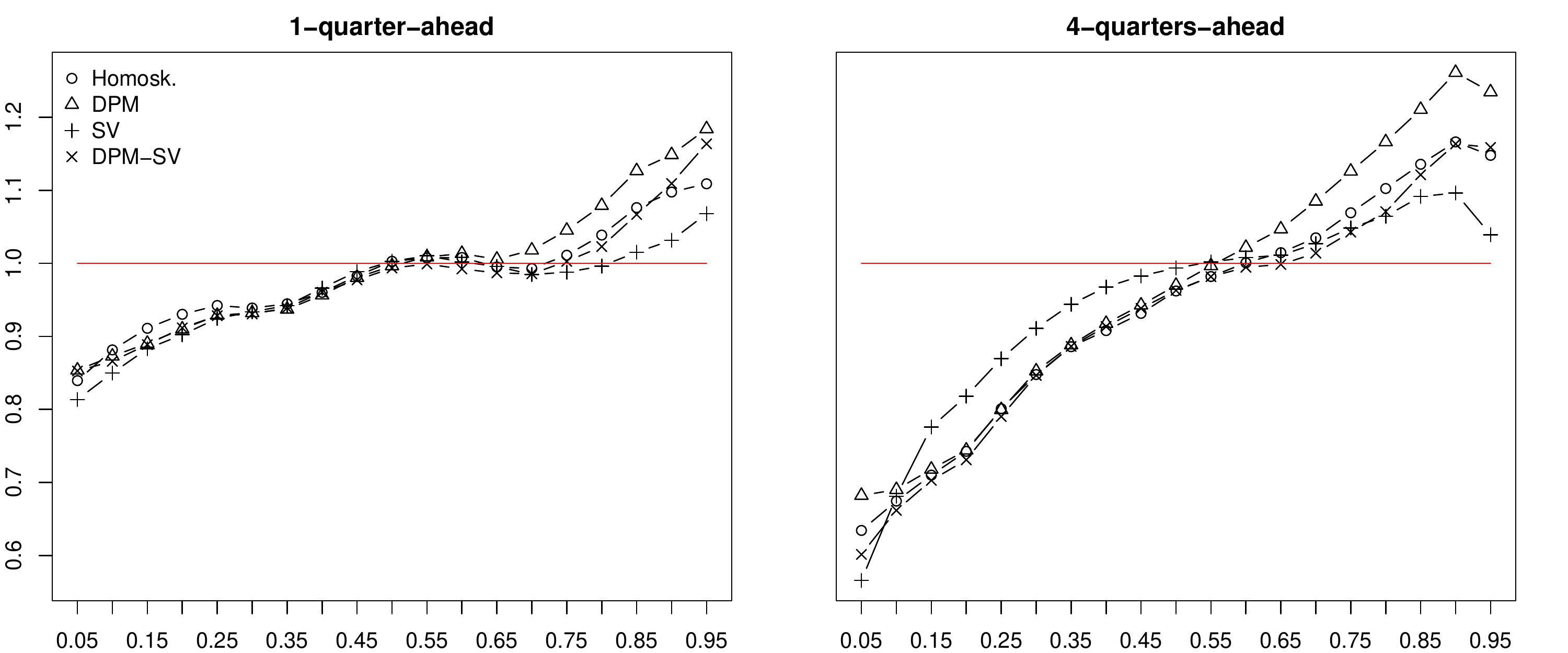}
   \end{minipage}
   \begin{minipage}[c]{\linewidth}\vspace{.45cm}
   \textbf{(b) Gaussian process with subspace shrinkage}
\end{minipage}
      \begin{minipage}[c]{\linewidth}
   \includegraphics[scale=0.58]{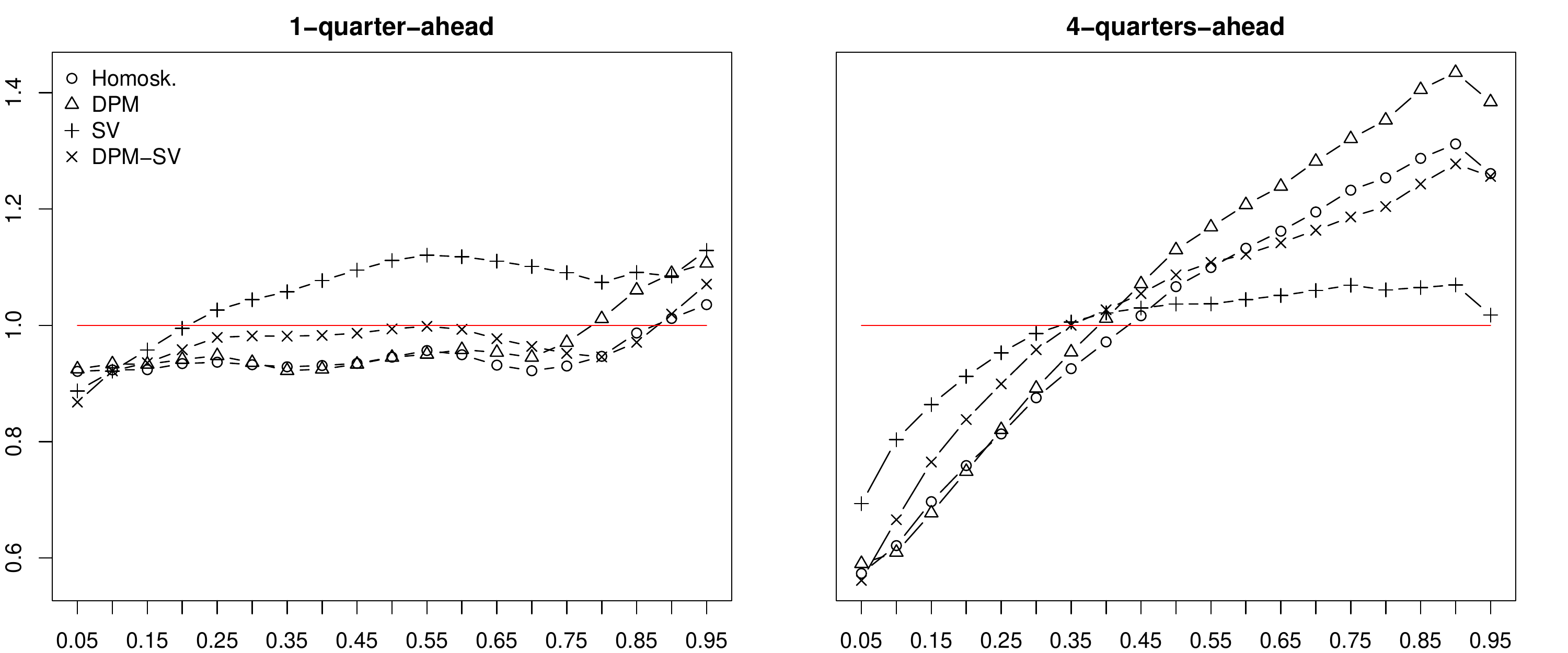}
   \end{minipage}
   \caption{Quantile scores across volatility specifications and forecast horizons: Moderately sized models}\label{fig: qs_small}
\end{figure}

\begin{figure}[t]
\begin{minipage}[c]{\linewidth}
   \textbf{(a) Gaussian process}
\end{minipage}
   \begin{minipage}[c]{\linewidth}
   \includegraphics[scale=0.58]{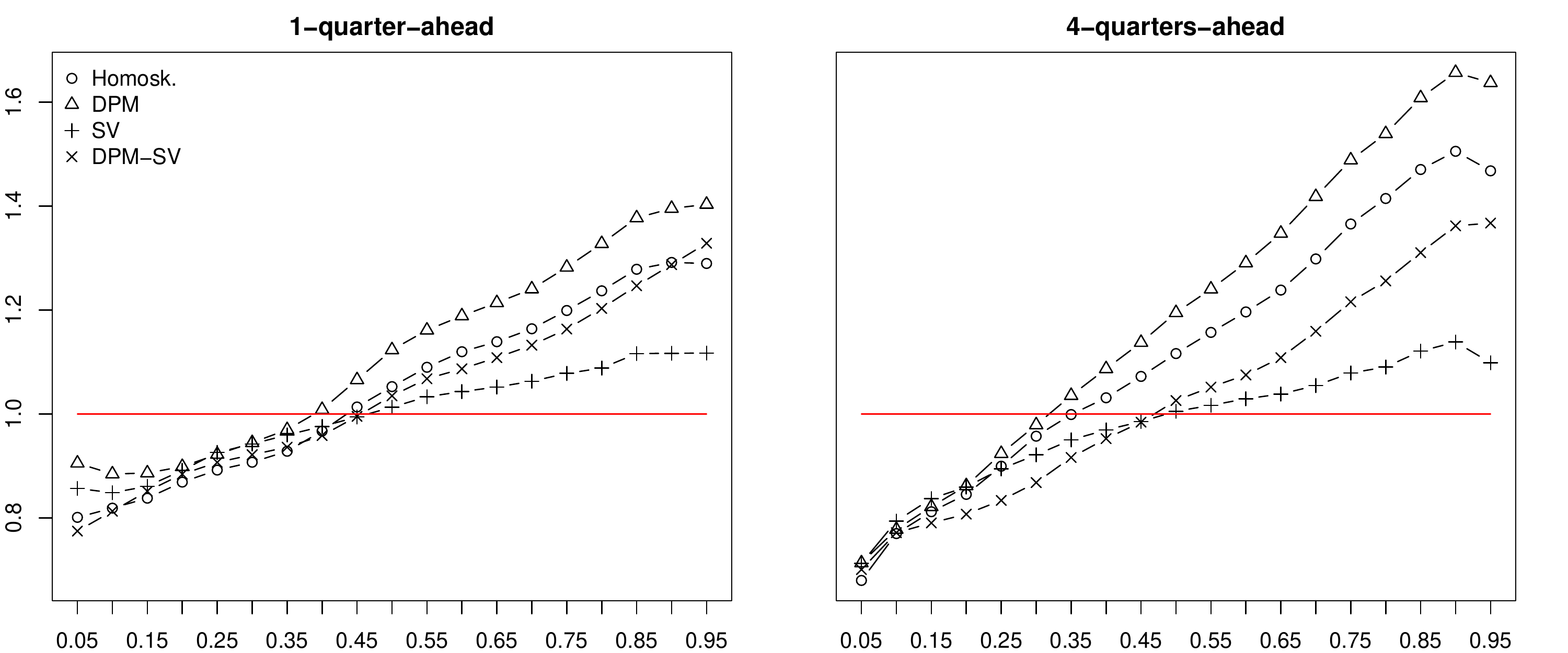}
   \end{minipage}
   \begin{minipage}[c]{\linewidth}\vspace{.45cm}
   \textbf{(b) Gaussian process with subspace shrinkage}
\end{minipage}
      \begin{minipage}[c]{\linewidth}
   \includegraphics[scale=0.58]{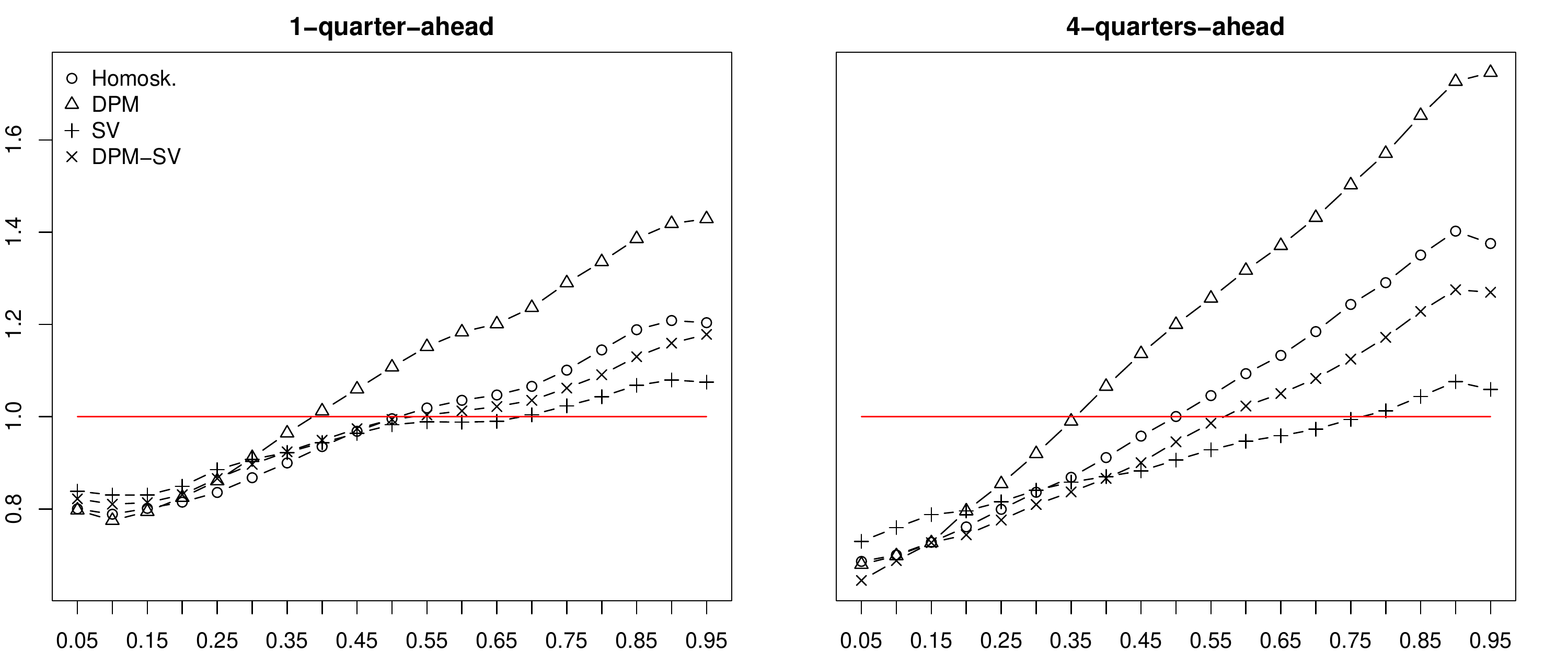}
   \end{minipage}
   \caption{Quantile scores across volatility specifications and forecast horizons: Large models}\label{fig: qs_large}
\end{figure}

To assess efficacy in forecasting tail risks to inflation --- both left and right tails --- Figures \ref{fig: qs_small} and \ref{fig: qs_large} present quantile scores for GP and GP-sub specifications with, respectively, moderately sized and large models.  In this comparison, too, scores are presented as relative to the UC-SV benchmark. The main message emerging from these figures is that the GP models are much better in the left tail (low inflation) than in the right tail (high inflation). More specifically, from Figure \ref{fig: qs_small}'s $h=1$ results with the moderately sized set of variables, in the left tail, GP is better than GP-sub, and both are 10-20 percent better than UC-SV.  At the 5 percent and 10 percent quantiles, the specifications  with SV and DPM-SV are most accurate.  With the moderately sized set of variables, the left-tail gains are noticeably larger for $h=4$ (reaching as much as 40 percent) than for $h=1$.  However, as the quantile considered moves toward the median and into the right tail, the score performance deteriorates.  Indeed, in the right tail, the nonparametric models are less accurate than the UC-SV baseline, by 10-20 percent for GP and a bit less for GP-sub (and as much as 40 percent at the four-quarters-ahead horizon). Results with large models are broadly similar to those with moderately sized models:  The patterns for large models in Figure \ref{fig: qs_large} and moderately sized models in Figure \ref{fig: qs_small} are quite similar.  Quantitatively, the gains in the left tail are sometimes more modest with large models than with moderately sized ones, whereas the losses in the right tail are sometimes larger with large models than with moderately sized ones.

These quantile scores are computed over the full hold-out period (i.e., 1980:Q1 to 2021:Q3) and might thus mask important temporal variation in tail forecasting performance. To understand whether the tail performance described above is stable over time, Table \ref{tab_qs} reports quantile scores (again, relative to the UC-SV model) for four subsamples:  1980-90, 1991-2000, 2001-10, and 2011-21. For $h=1$, the gains in the left tail are present in all periods except 1991-2000, and they are often greater for large-scale GP than for moderately sized GP.  Given the GP specification, there is no clear-cut ranking of the error process.  For $h=4$ and left-tail forecasts, the moderately sized GP-sub is better than UC-SV in most periods and for most error specifications.  In the 1991-2000 period that stands as the exception, in most cases the benchmark UC-SV model yields better scores compared to the GP models.  Consistent with the full sample results, in the three subsamples before 2011, in right-tail quantile scores most models are beaten by UC-SV, sometimes with large losses.  But for the 2011-21 period, patterns reverse, and most GP models beat the forecast accuracy of UC-SV in the right tail. In particular, in this period the large GP and GP-sub with SV specifications are better than UC-SV for both left and right tails, with gains of up to 5 percent in the former and 44 percent in the latter (notice the reversal with respect to $h=1$). Overall, though the full sample performance in the right tail is disappointing, large-scale GP with SV behaves well both in the left and in the right tail in the most recent period, with systematically better quantile scores than UC-SV. 

In the empirical appendix (see Figures \ref{fig:cumQS_small} and \ref{fig:cumQS_large}) we also show how relative cumulative quantile scores (for $p=0.05$ and $p=0.95$) evolve over the hold-out period. These figures tell a story similar to the one provided above. In particular, in the high-inflation periods (i.e., the early 1980s and 2021) our set of nonparametric models also does well in the right tail.

%Instead, in line with full sample results, for $h=1$ and right tail, the performance generally worsens, but small GP-sub fares well in 1980-90 and 2011-21. Moreover, for $h=1$, in 2011-2021 large GP and GP-sub are better than UC-SV for both left and right tail, with gains up to 30\% in the former and 8\% in the latter. For $h=4$ and left tail, small GP-sub is better than UC-SV in most periods and for most error specifications. For the right tail, up to 2010 most models are beaten by UC-SV, sometimes with large losses.  This pattern may be traceable to the low and stable inflation that prevailed from the early 1990s until the Great Recession.  Yet, in the most recent period, 2011-21, there is a reversal, and most GP models beat the UC-SV also in the right tail. In particular, in this period large GP and GP-sub with SV are better than UC-SV for both left and right tails, with gains up to 5\% in the former and 44\% in the latter (notice inversion with respect to $h=1$). Overall, though the full sample performance in the right tail is disappointing, large GP with SV behaves well both in the left and in the right tail in the most recent period, with systematically better quantile scores than UC-SV.

\begin{table}[t]
\centering
\scalebox{0.6}{
\begin{tabular}{lllllllllllllllllllll}
  \hline
  Subsample &\multicolumn{4}{r}{1980 -- 1990}&&\multicolumn{4}{r}{1991 -- 2000} && \multicolumn{4}{r}{2001 -- 2010}&& \multicolumn{4}{r}{2011 -- 2021} & \\
      \multicolumn{21}{c}{}\\ 
    \multicolumn{21}{c}{Moderately sized models}\\ 
  \multicolumn{21}{l}{Gaussian process}\\ \midrule
   &  & Homosk. & DPM & SV & DPM-SV &  & Homosk. & DPM & SV & DPM-SV &  & Homosk. & DPM & SV & DPM-SV &  & Homosk. & DPM & SV & DPM-SV \\ 
     $h=1$ &  &  &  &  &  &  &  &  &  &  &  &  &  &  &  &  &  &  &  &  \\ 
  0.05 &  & 0.72 & 0.76 & 0.69 & 0.67 &  & 1.27 & 1.32 & 1.09 & 1.29 &  & 0.81 & 0.83 & 0.83 & 0.84 &  & 0.83 & 0.8 & 0.79 & 0.87 \\ 
  0.1 &  & 0.7 & 0.7 & 0.76 & 0.75 &  & 1.27 & 1.25 & 1.12 & 1.2 &  & 0.91 & 0.91 & 0.86 & 0.87 &  & 0.91 & 0.87 & 0.84 & 0.89 \\ 
  0.9 &  & 1.05 & 1.14 & 0.84 & 1 &  & 1.42 & 1.54 & 1.16 & 1.44 &  & 1.06 & 1.03 & 1.19 & 1.12 &  & 0.99 & 1.06 & 0.98 & 1.01 \\ 
  0.95 &  & 1.17 & 1.27 & 0.95 & 1.17 &  & 1.38 & 1.49 & 1.13 & 1.48 &  & 0.93 & 0.96 & 1.06 & 0.99 &  & 1.1 & 1.18 & 1.16 & 1.17 \\ 
  $h=4$ &  &  &  &  &  &  &  &  &  &  &  &  &  &  &  &  &  &  &  &  \\ 
 0.05.1 &  & 0.5 & 0.55 & 0.54 & 0.58 &  & 1.05 & 1.12 & 0.93 & 1 &  & 0.46 & 0.49 & 0.38 & 0.39 &  & 1.19 & 1.28 & 0.94 & 1.03 \\ 
  0.1 &  & 0.52 & 0.49 & 0.64 & 0.57 &  & 1 & 1.07 & 1.06 & 1.1 &  & 0.57 & 0.58 & 0.5 & 0.5 &  & 1.04 & 1.11 & 0.95 & 0.94 \\ 
  0.9 &  & 1.32 & 1.39 & 1.08 & 1.25 &  & 1.58 & 1.74 & 1.22 & 1.48 &  & 1.15 & 1.26 & 1.41 & 1.25 &  & 0.73 & 0.81 & 0.77 & 0.79 \\ 
  0.95 &  & 1.37 & 1.45 & 1.13 & 1.33 &  & 1.69 & 1.85 & 1.28 & 1.66 &  & 1.12 & 1.19 & 1.34 & 1.25 &  & 0.67 & 0.74 & 0.59 & 0.66 \\ 
   &  &  &  &  &  &  &  &  &  &  &  &  &  &  &  &  &  &  &  &  \\ 
   \multicolumn{21}{l}{Gaussian process with subspace shrinkage}\\ \midrule
        $h=1$ &  &  &  &  &  &  &  &  &  &  &  &  &  &  &  &  &  &  &  &  \\ 
  0.05 &  & 0.79 & 0.8 & 0.79 & 0.78 &  & 1.32 & 1.4 & 1.43 & 1.43 &  & 0.88 & 0.87 & 0.77 & 0.74 &  & 0.99 & 0.98 & 1.01 & 0.99 \\ 
  0.1 &  & 0.76 & 0.79 & 0.8 & 0.79 &  & 1.29 & 1.35 & 1.5 & 1.39 &  & 0.93 & 0.94 & 0.85 & 0.88 &  & 0.98 & 0.96 & 0.98 & 0.98 \\ 
  0.9 &  & 0.95 & 1.01 & 0.86 & 0.82 &  & 1.27 & 1.5 & 1.12 & 1.12 &  & 1.02 & 1.03 & 1.48 & 1.22 &  & 0.92 & 1 & 0.89 & 0.96 \\ 
  0.95 &  & 1.05 & 1.13 & 0.87 & 0.93 &  & 1.25 & 1.47 & 0.92 & 1.19 &  & 0.94 & 0.92 & 1.6 & 1.12 &  & 1.01 & 1.09 & 1.02 & 1.1 \\ 
  $h=4$ &  &  &  &  &  &  &  &  &  &  &  &  &  &  &  &  &  &  &  &  \\ 
 0.05 &  & 0.43 & 0.44 & 0.45 & 0.37 &  & 0.74 & 0.84 & 1 & 0.74 &  & 0.55 & 0.53 & 0.56 & 0.52 &  & 0.89 & 0.95 & 1.48 & 1.02 \\ 
  0.1 &  & 0.42 & 0.43 & 0.63 & 0.53 &  & 0.84 & 0.86 & 1.1 & 0.89 &  & 0.64 & 0.61 & 0.68 & 0.6 &  & 0.86 & 0.82 & 1.29 & 1 \\ 
  0.9 &  & 1.35 & 1.43 & 1.06 & 1.25 &  & 1.88 & 2.08 & 1.29 & 1.79 &  & 1.29 & 1.4 & 1.29 & 1.34 &  & 0.92 & 1.05 & 0.75 & 0.93 \\ 
  0.95 &  & 1.38 & 1.47 & 1.09 & 1.31 &  & 1.94 & 2.14 & 1.32 & 1.91 &  & 1.24 & 1.36 & 1.22 & 1.32 &  & 0.8 & 0.92 & 0.64 & 0.81 \\ 
  \multicolumn{21}{c}{Large-scale models}\\ 
 \multicolumn{21}{l}{Gaussian process}\\ \midrule
  $h=1$ &  &  &  &  &  &  &  &  &  &  &  &  &  &  &  &  &  &  &  &  \\ 
  0.05 &  & 0.65 & 0.73 & 0.68 & 0.63 &  & 1.24 & 1.3 & 1.24 & 1.25 &  & 0.82 & 0.99 & 0.93 & 0.76 &  & 0.72 & 0.72 & 0.69 & 0.74 \\ 
  0.1 &  & 0.58 & 0.62 & 0.69 & 0.64 &  & 1.21 & 1.25 & 1.14 & 1.19 &  & 0.92 & 1.04 & 0.95 & 0.87 &  & 0.77 & 0.78 & 0.75 & 0.77 \\ 
  0.9 &  & 1.47 & 1.57 & 1.24 & 1.44 &  & 1.97 & 2.1 & 1.31 & 1.85 &  & 1.05 & 1.17 & 1.08 & 1.11 &  & 0.97 & 1.05 & 0.92 & 1.01 \\ 
  0.95 &  & 1.58 & 1.72 & 1.34 & 1.58 &  & 1.85 & 2 & 1.27 & 1.83 &  & 1.05 & 1.17 & 1.01 & 1.13 &  & 0.96 & 1.01 & 0.96 & 1.03 \\ 
  $h=4$ &  &  &  &  &  &  &  &  &  &  &  &  &  &  &  &  &  &  &  &  \\ 
  0.05 &  & 0.54 & 0.52 & 0.67 & 0.57 &  & 1.02 & 1.09 & 1.12 & 1.05 &  & 0.51 & 0.59 & 0.57 & 0.57 &  & 1.32 & 1.28 & 0.95 & 1.19 \\ 
  0.1 &  & 0.65 & 0.62 & 0.76 & 0.67 &  & 0.93 & 0.96 & 1.12 & 1.02 &  & 0.69 & 0.76 & 0.65 & 0.66 &  & 1.15 & 1.07 & 0.99 & 1.12 \\ 
  0.9 &  & 1.76 & 1.87 & 1.55 & 1.7 &  & 2.19 & 2.31 & 1.37 & 1.81 &  & 1.37 & 1.6 & 1.03 & 1.21 &  & 0.9 & 1.05 & 0.63 & 0.83 \\ 
  0.95 &  & 1.77 & 1.92 & 1.57 & 1.75 &  & 2.27 & 2.45 & 1.42 & 1.95 &  & 1.38 & 1.61 & 1 & 1.28 &  & 0.81 & 0.96 & 0.56 & 0.77 \\ 
   \multicolumn{21}{l}{Gaussian process with subspace shrinkage}\\ \midrule
  $h=1$ &  &  &  &  &  &  &  &  &  &  &  &  &  &  &  &  &  &  &  &  \\ 
  0.05 &  & 0.66 & 0.67 & 0.6 & 0.65 &  & 1.1 & 1.08 & 1.04 & 0.93 &  & 0.85 & 0.83 & 0.95 & 0.92 &  & 0.69 & 0.73 & 0.74 & 0.72 \\ 
  0.1 &  & 0.58 & 0.58 & 0.6 & 0.58 &  & 1.09 & 1.03 & 1 & 0.97 &  & 0.88 & 0.85 & 0.96 & 0.93 &  & 0.75 & 0.78 & 0.82 & 0.81 \\ 
  0.9 &  & 1.24 & 1.57 & 1.08 & 1.16 &  & 1.93 & 2.34 & 1.45 & 1.67 &  & 0.93 & 1.08 & 0.97 & 0.98 &  & 1.01 & 1.06 & 0.94 & 1 \\ 
  0.95 &  & 1.35 & 1.73 & 1.19 & 1.3 &  & 1.8 & 2.19 & 1.36 & 1.58 &  & 0.94 & 1.1 & 0.92 & 0.97 &  & 1.01 & 1.08 & 0.99 & 1.07 \\ 
  $h=4$ &  &  &  &  &  &  &  &  &  &  &  &  &  &  &  &  &  &  &  &  \\ 
  0.05 &  & 0.53 & 0.52 & 0.49 & 0.55 &  & 0.92 & 0.86 & 1.14 & 0.74 &  & 0.61 & 0.62 & 0.74 & 0.65 &  & 1.15 & 1.15 & 0.88 & 0.81 \\ 
  0.1 &  & 0.53 & 0.59 & 0.51 & 0.53 &  & 0.83 & 0.73 & 1.12 & 0.9 &  & 0.72 & 0.71 & 0.77 & 0.72 &  & 0.96 & 0.93 & 1 & 0.79 \\ 
  0.9&  & 1.47 & 1.91 & 1.25 & 1.4 &  & 2.18 & 2.66 & 1.39 & 1.81 &  & 1.22 & 1.43 & 1.01 & 1.17 &  & 0.98 & 1.16 & 0.74 & 0.87 \\ 
  0.95 &  & 1.52 & 1.99 & 1.32 & 1.49 &  & 2.26 & 2.83 & 1.44 & 1.91 &  & 1.24 & 1.55 & 1.02 & 1.18 &  & 0.87 & 1.08 & 0.65 & 0.78 \\ 
   \hline
\end{tabular}
}
\caption{Quantile scores relative to the UC-SV model. Averages across subsamples.}\label{tab_qs}
\end{table}

\subsection{Model calibration}  % : Probability integral transforms

\begin{figure}[t!]
    \centering
    \begin{minipage}[c]{\linewidth}
   \textbf{(a) Moderately sized models}
\end{minipage}
    \includegraphics[scale=0.65]{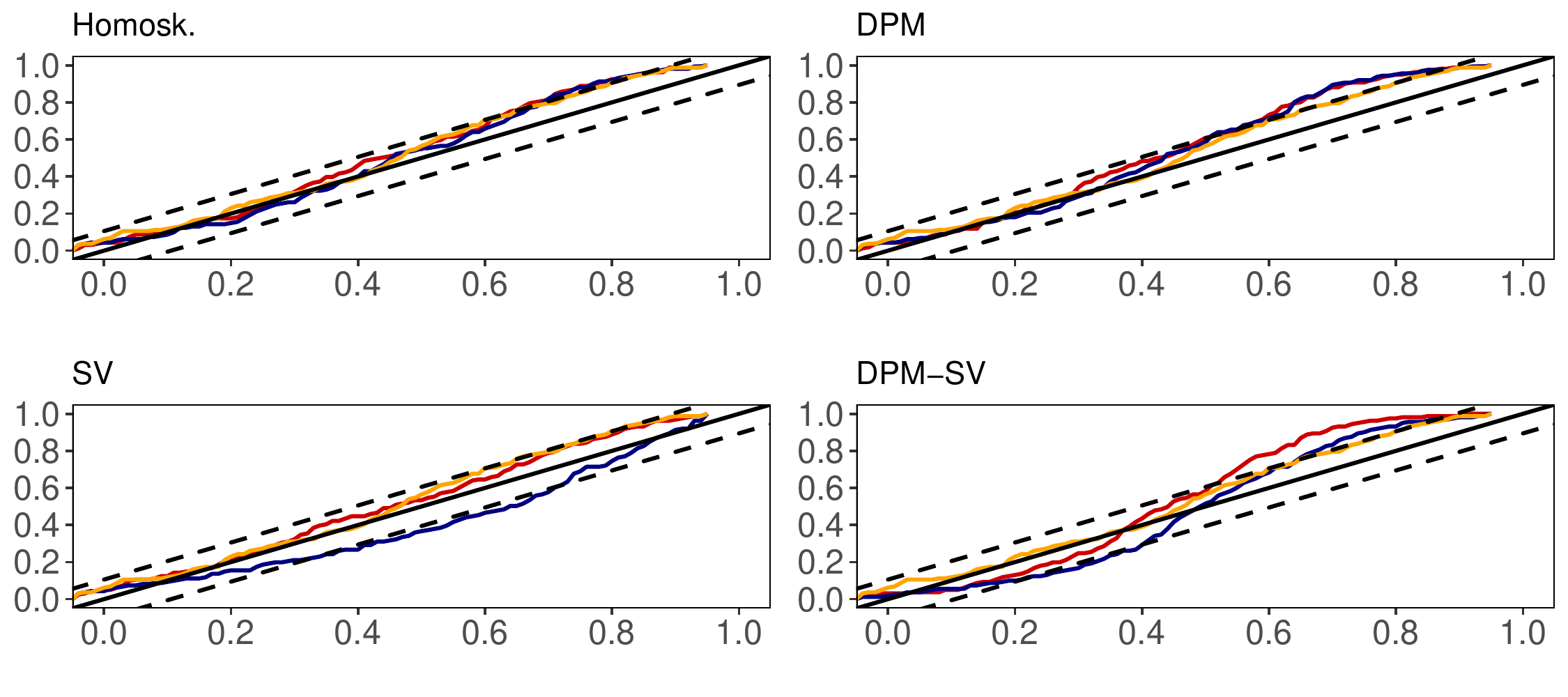}
     \begin{minipage}[c]{\linewidth}
   \textbf{(b) Large-scale models}
\end{minipage}
    \includegraphics[scale=0.65]{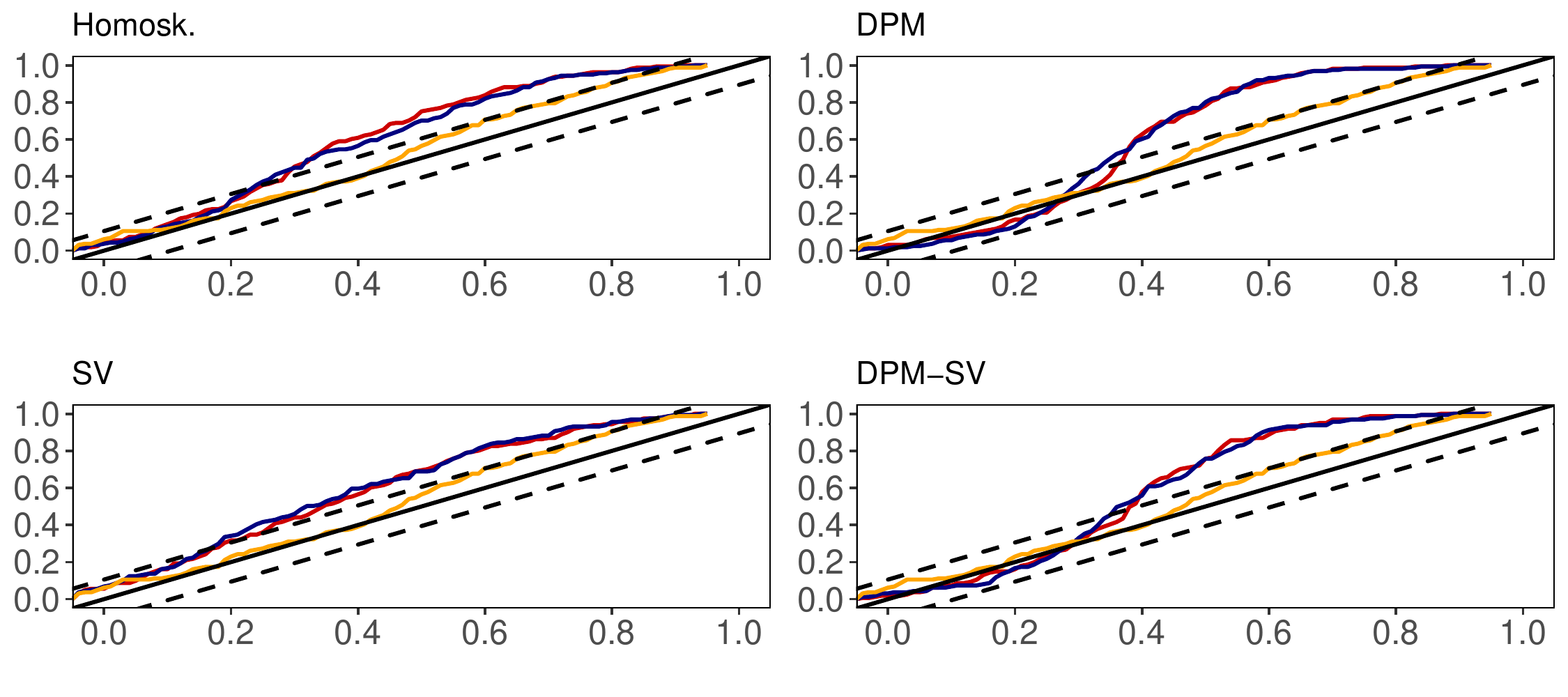}
    \caption*{\footnotesize \textbf{Notes}: The red line refers to the Gaussian process regression, the blue line to the GP regression with subspace shrinkage, and the orange line is the UC-SV model.}
    \caption{\cite{rossi2019alternative} diagnostic plots for the one-quarter-ahead predictive densities.}
    \label{fig:PIT_one}
\end{figure}

\begin{figure}[hbt!]
    \centering
    \begin{minipage}[c]{\linewidth}
   \textbf{(a) Moderately sized models}
\end{minipage}
    \includegraphics[scale=0.65]{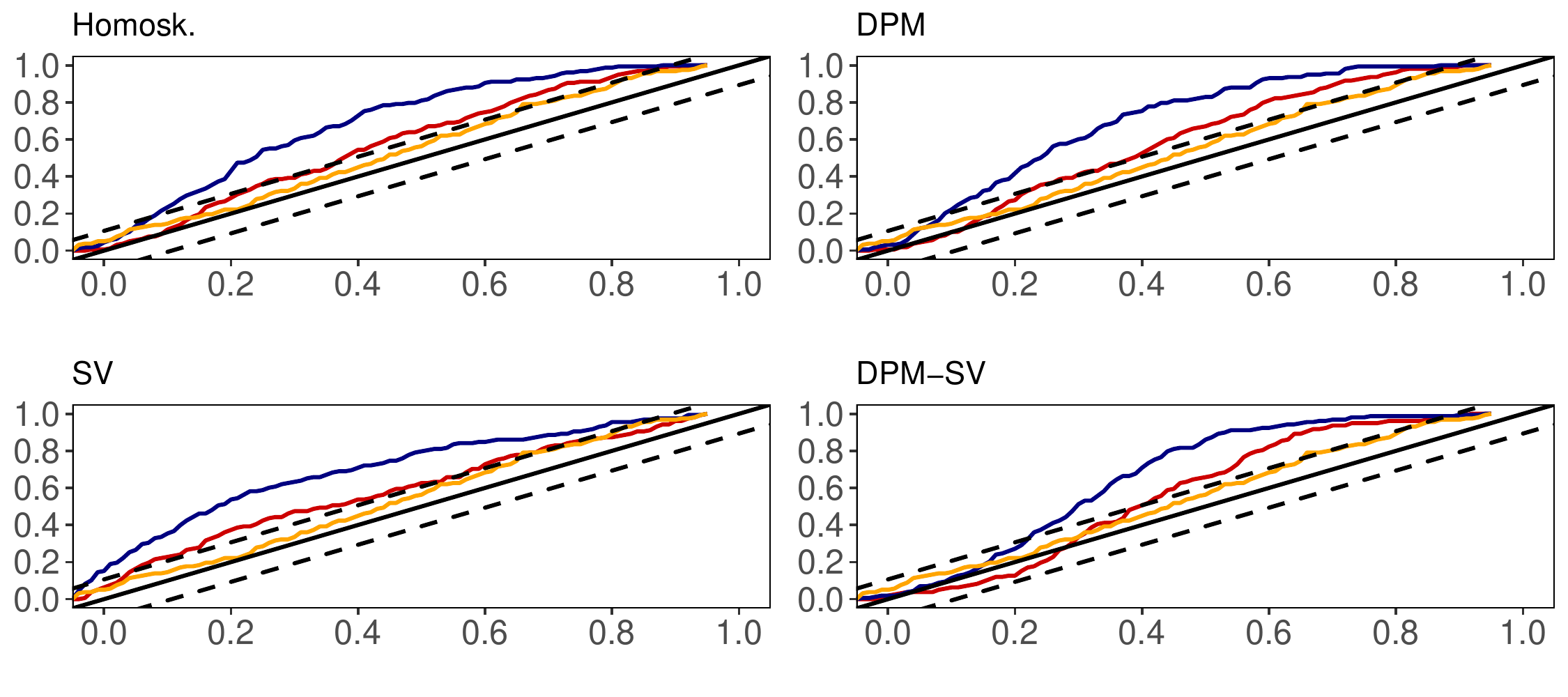}
     \begin{minipage}[c]{\linewidth}
   \textbf{(b) Large-scale models}
\end{minipage}
    \includegraphics[scale=0.65]{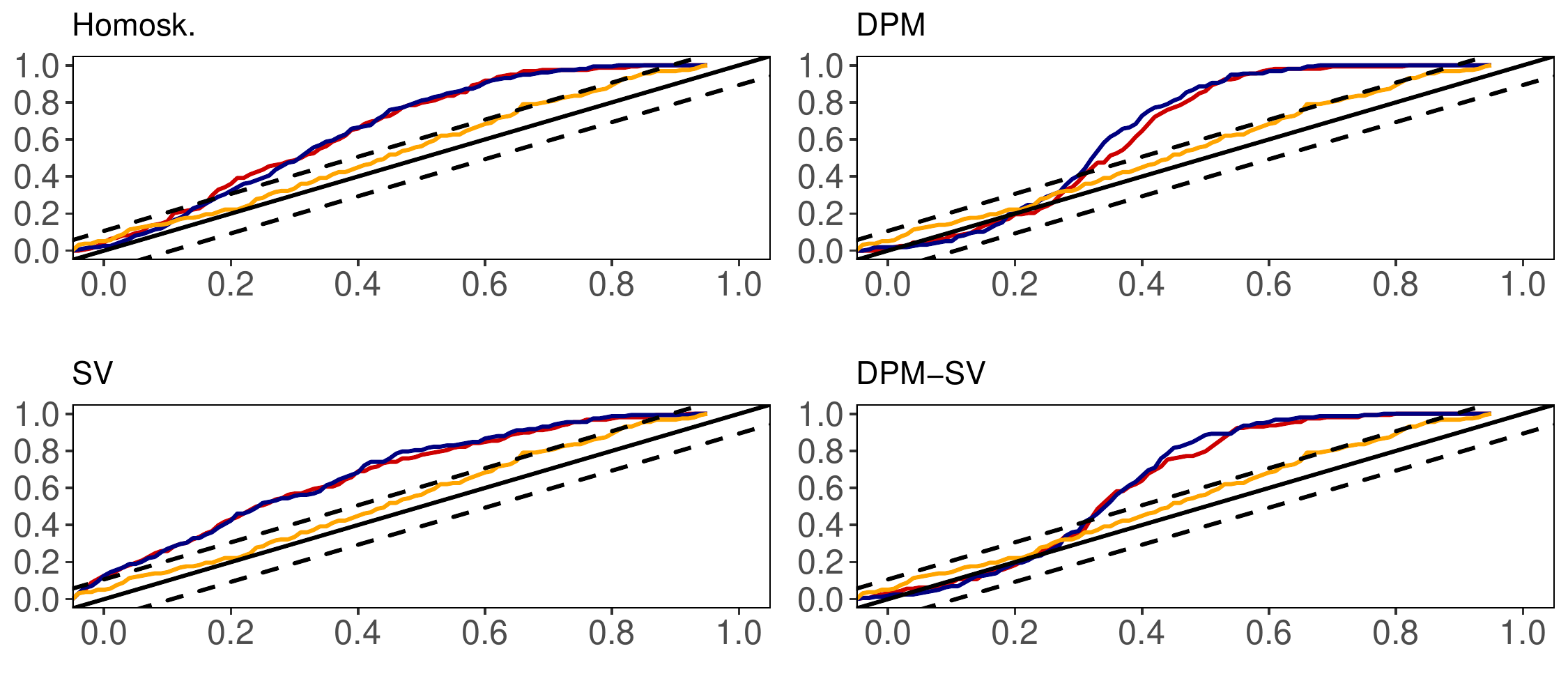}
    \caption*{\footnotesize \textbf{Notes}: The red line refers to the Gaussian process regression, the blue line to the GP regression with subspace shrinkage, and the orange line is the UC-SV model.}
    \caption{\cite{rossi2019alternative} diagnostic plots for the four-quarters-ahead predictive densities.}
    \label{fig:PIT_four}
\end{figure}

The previous results have been based on relative model performance. To gauge how well the corresponding predictive densities are calibrated, 
Figures \ref{fig:PIT_one} and \ref{fig:PIT_four} report the \cite{rossi2019alternative} diagnostic plots for, respectively, the one-step- and four-steps-ahead predictive densities over the 1980-2021 evaluation sample. These diagnostics provide information about whether the predictive densities are correctly specified given the model and estimation technique specified by the researcher. They are essentially QQ-plots that allow us to directly investigate in what region of the predictive distribution the mis-calibration occurs. In these figures we compare distributions for the UC-SV and GP models, including GP results under our four different error specifications.  In each plot, the solid black and dotted lines represent the benchmark of correct specification with confidence bands.

These results indicate that the predictive densities of the GP-SV and UC-SV specifications are closest to being correctly specified, whereas the densities of other models display some noticeable mis-specification, likely related to the bad predictive performance in the right tail noted above (unreported probability integral transforms yield similar patterns and conclusions).  For both horizons, the UC-SV model's quantiles of the predictive density lay within the correct-specification bands.  With the moderately sized set of variables, at the one-step-ahead horizon the quantiles of the predictive density also lay within the confidence band for the homoskedastic and SV versions of the GP model.  But significant departures from correct specification occur if the version of the GP model is changed, the forecast horizon is increased to 4, or the variable set is changed to the large-scale set.  In calibration, as in the quantile score performance presented above, the problems seem to be in the right tail of the distribution:  Departures from the correct specification line become more prevalent and larger as the quantile moves from the left to the right tail.

\subsection{A deeper look at the predictive densities over time}
\begin{figure}
    \centering
    \includegraphics[scale=0.5]{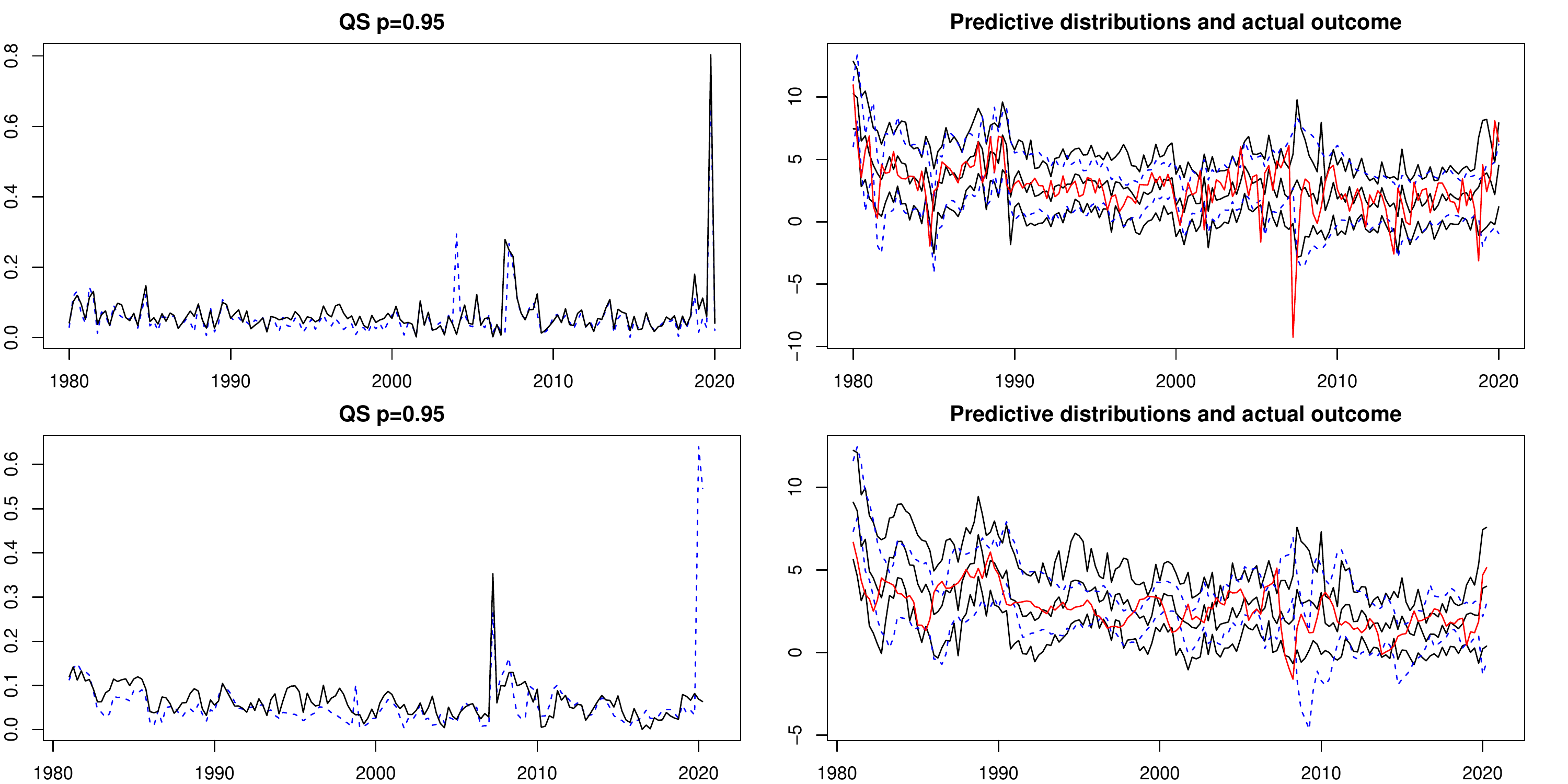}
        \caption*{\footnotesize \textbf{Notes}: Black solid lines refer to the GP regression and the DPM-SV specification on the shocks, while dashed gray lines denote the findings of the UC-SV model. The top panel refers to the one-quarter-ahead predictions, while the bottom panel shows the results for four-quarters-ahead.}
    \caption{Quantile scores ($p=0.95$) and $10, 50$ and $90^{th}$ percentiles of the one-step-ahead predictive distribution for the moderately sized data set}
    \label{fig:qs_preds}
\end{figure}

This sub-section focuses on the qualitative properties of the predictive distributions. Figure \ref{fig:qs_preds} graphs the evolution of right-tail quantile scores ($p$ = 0.95) and the $10, 50$ and $90^{th}$ percentiles of the predictive distribution.  In the interest of brevity, we report results for the selected specifications of the UC-SV benchmark and the GP model with DPM-SV for the moderately sized data set, for $h=1$ in the upper panel and $h=4$ in the lower panel.  

As shown in the panels in the left side of the figure, in most periods, the right-tail quantile scores are lower (better) for UC-SV than for GP, but in episodes in the mid-2000s and during the pandemic, the GP specification beats UC-SV.  As shown in the panels in the right side of the figure, the width between the $10$ and $90^{th}$ percentiles of the predictive distribution is generally narrower for UC-SV than for GP-sub, indicating a more concentrated predictive distribution for the former.  While the bands from the UC-SV model appear to move together and preserve symmetry in the predictive density, the bands from GP sometimes display asymmetry.  Most noticeably, around the 2008-2009 period, the model's 10th percentile drops more than does the 90$^{th}$ percentile.  As intended, the nonparametric specification appears to have more ability to capture asymmetries in predictive distributions.  Finally, it appears that the GP model more accurately predicts the rise of inflation during the pandemic. 

\begin{figure}[t!]
    \centering
    \begin{minipage}[t]{0.46\textwidth}
        \includegraphics[page=109, scale=0.3]{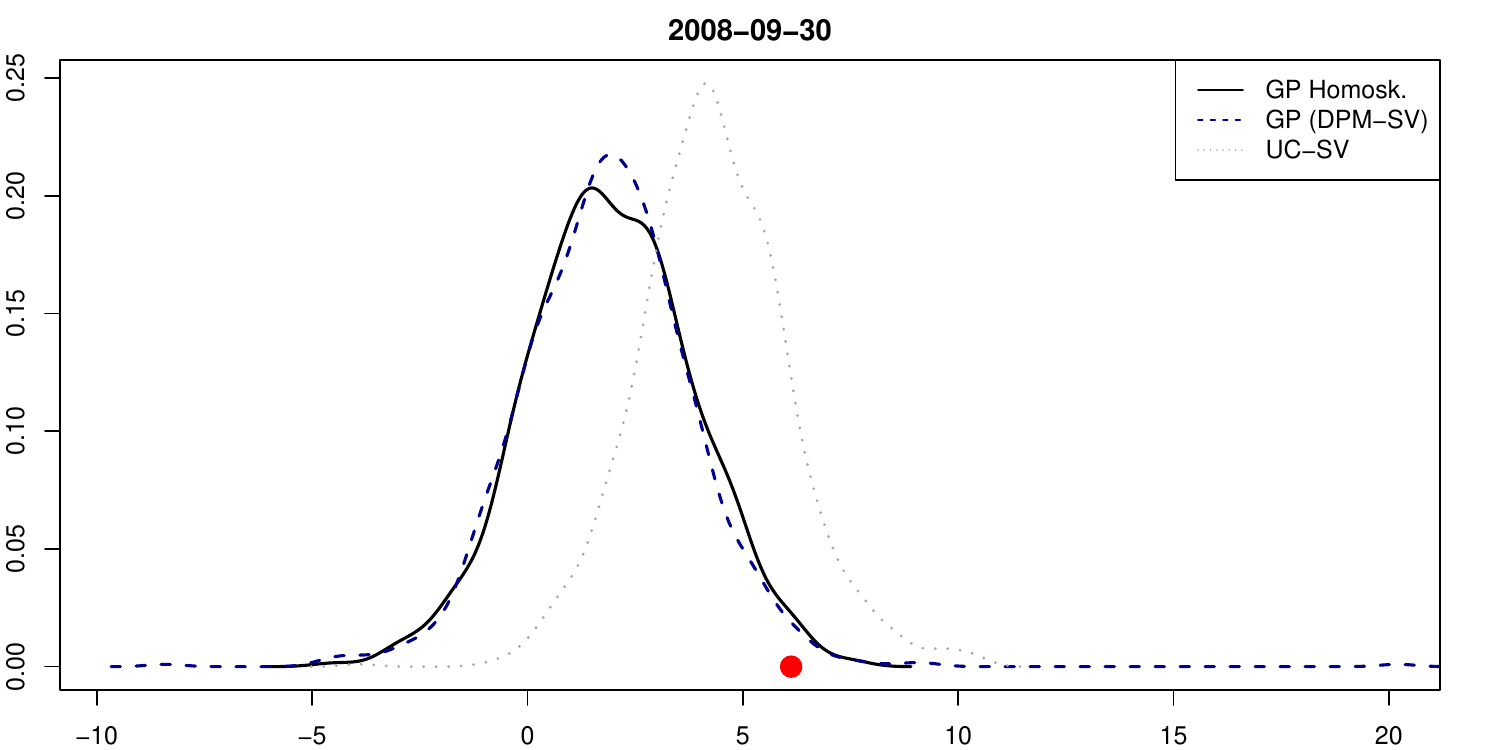}
    \includegraphics[page=110, scale=0.3]{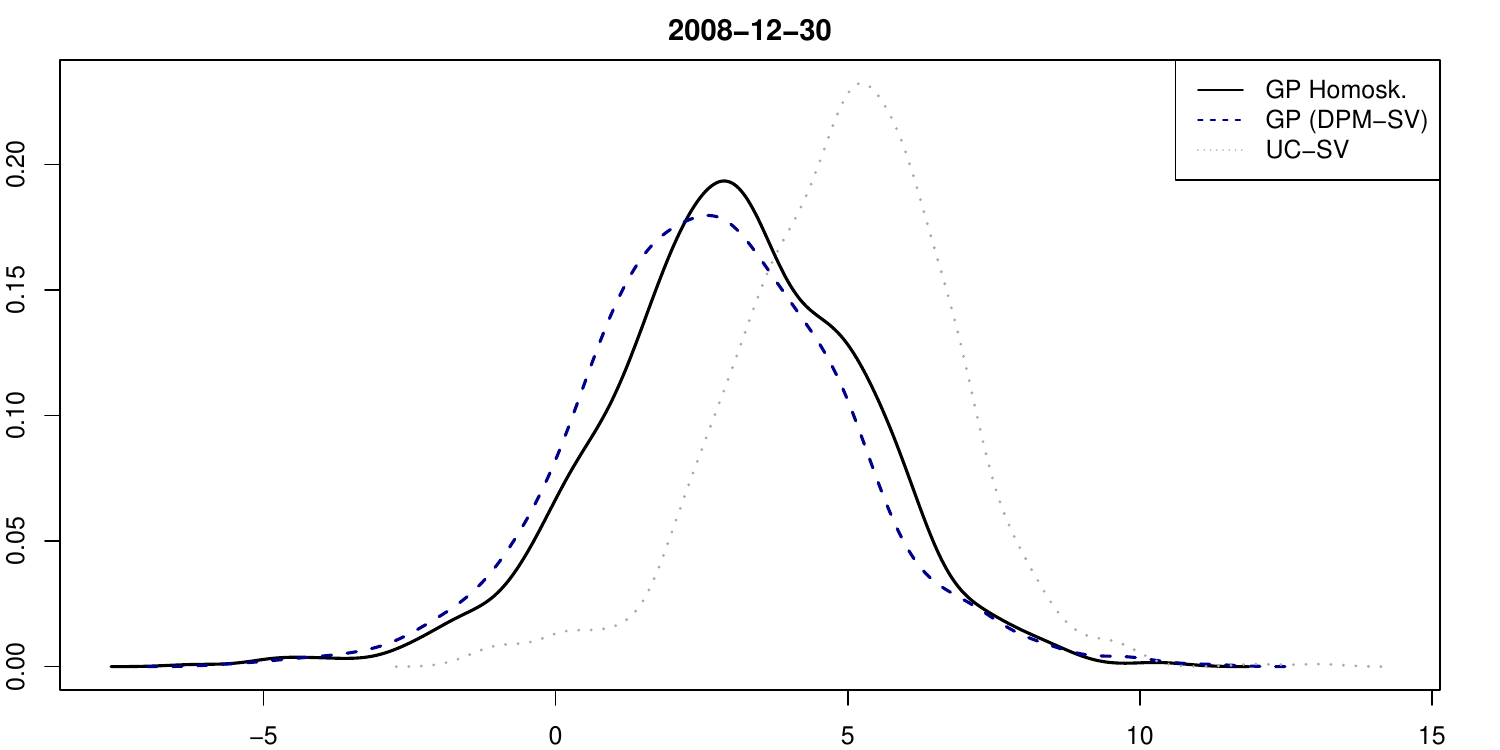}
    \includegraphics[page=111, scale=0.3]{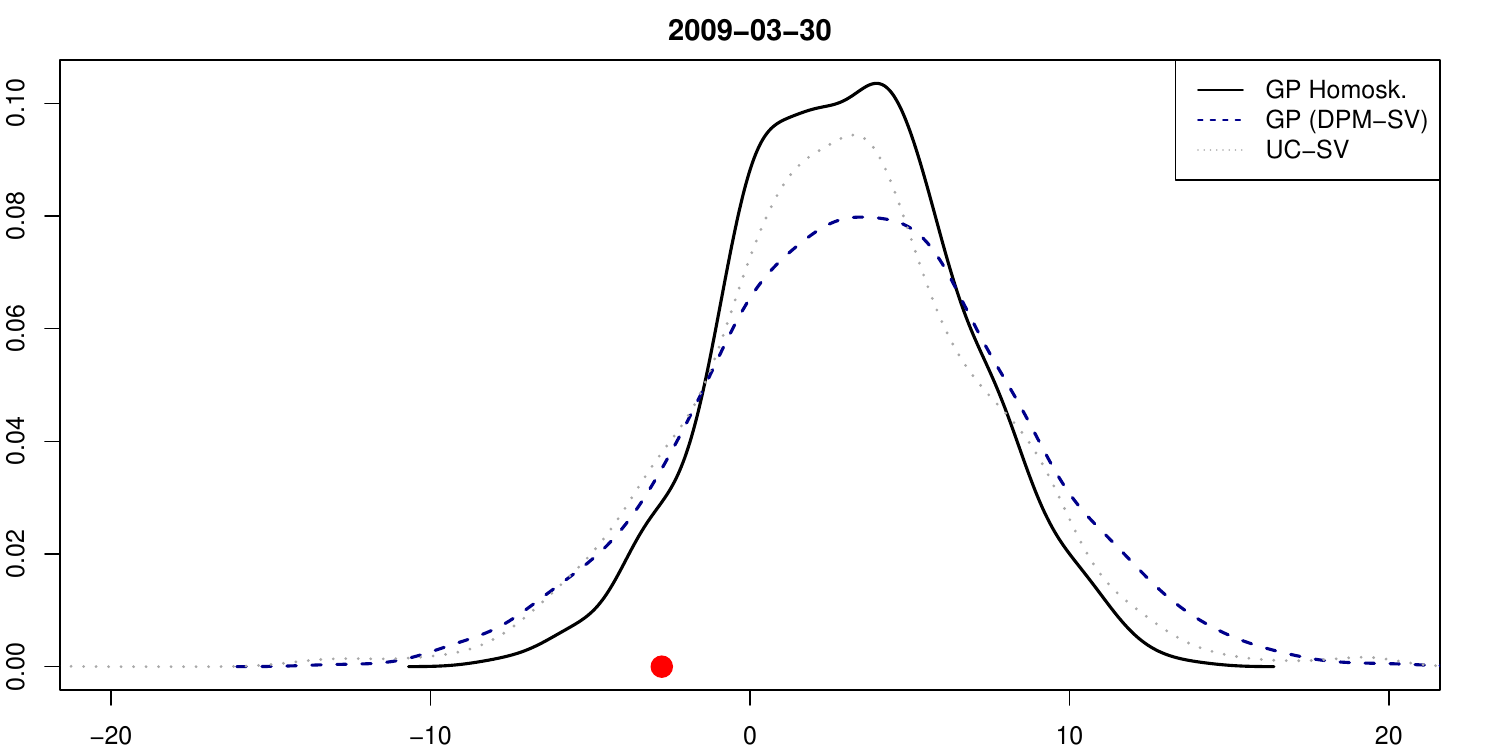}
    \includegraphics[page=112, scale=0.3]{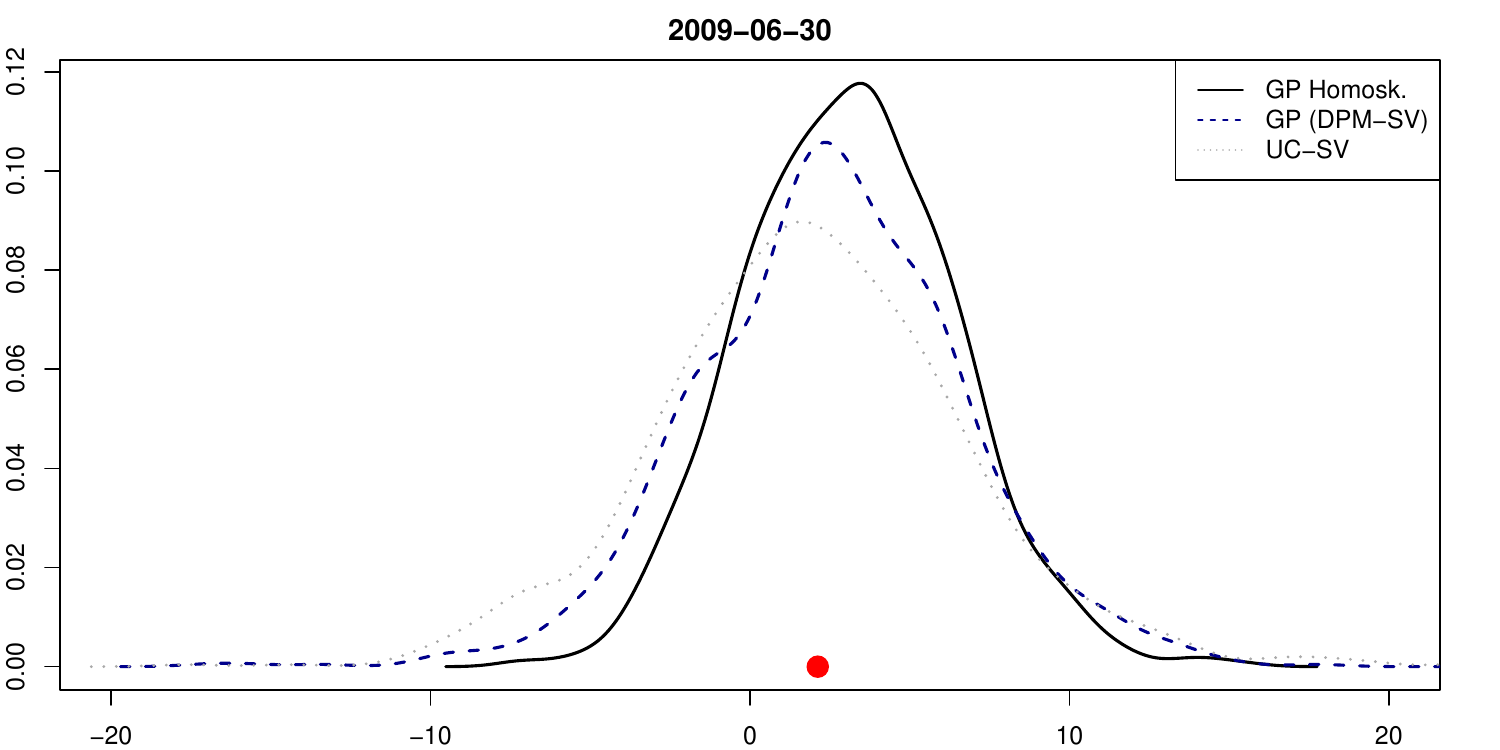}
    \end{minipage}
        \begin{minipage}[t]{0.47\textwidth}
        \includegraphics[page=155, scale=0.3]{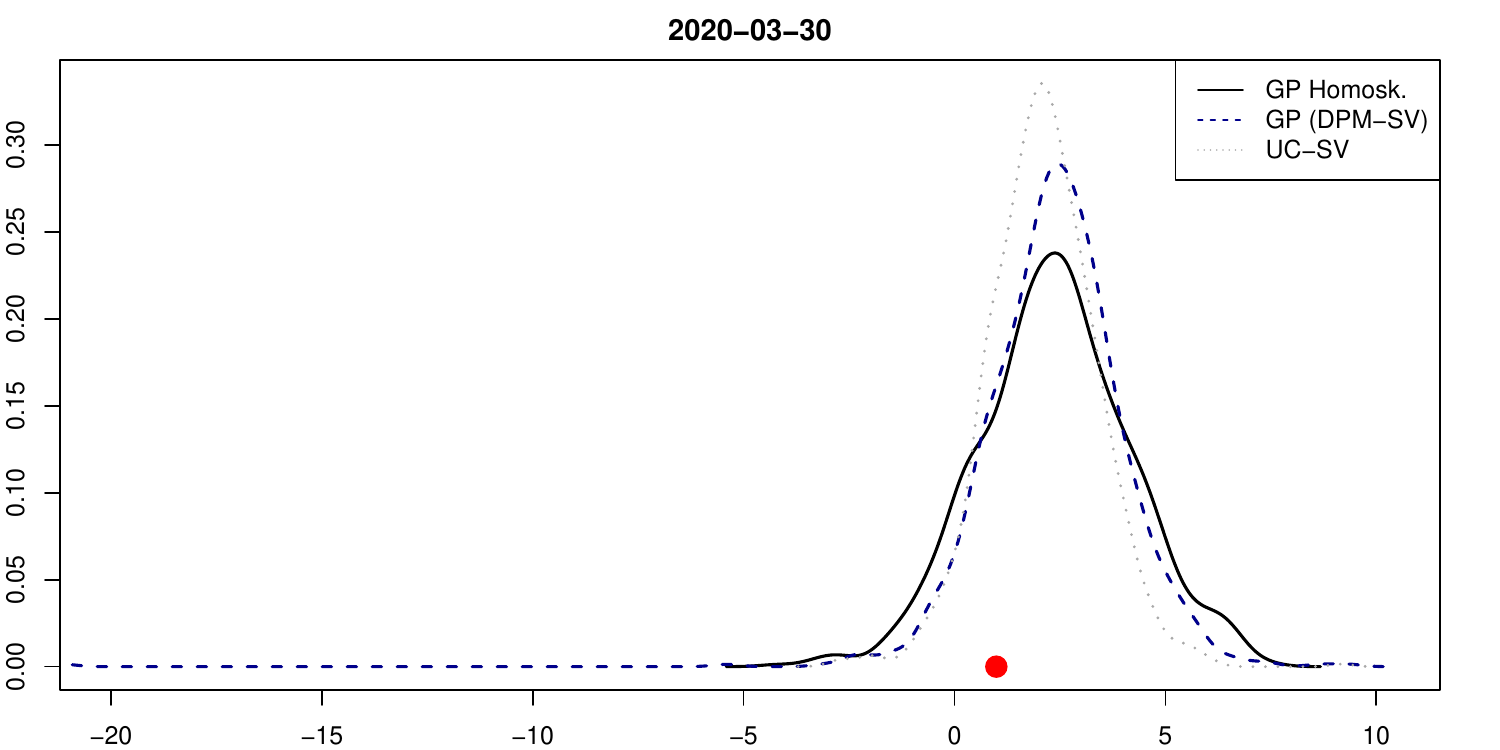}
    \includegraphics[page=156, scale=0.3]{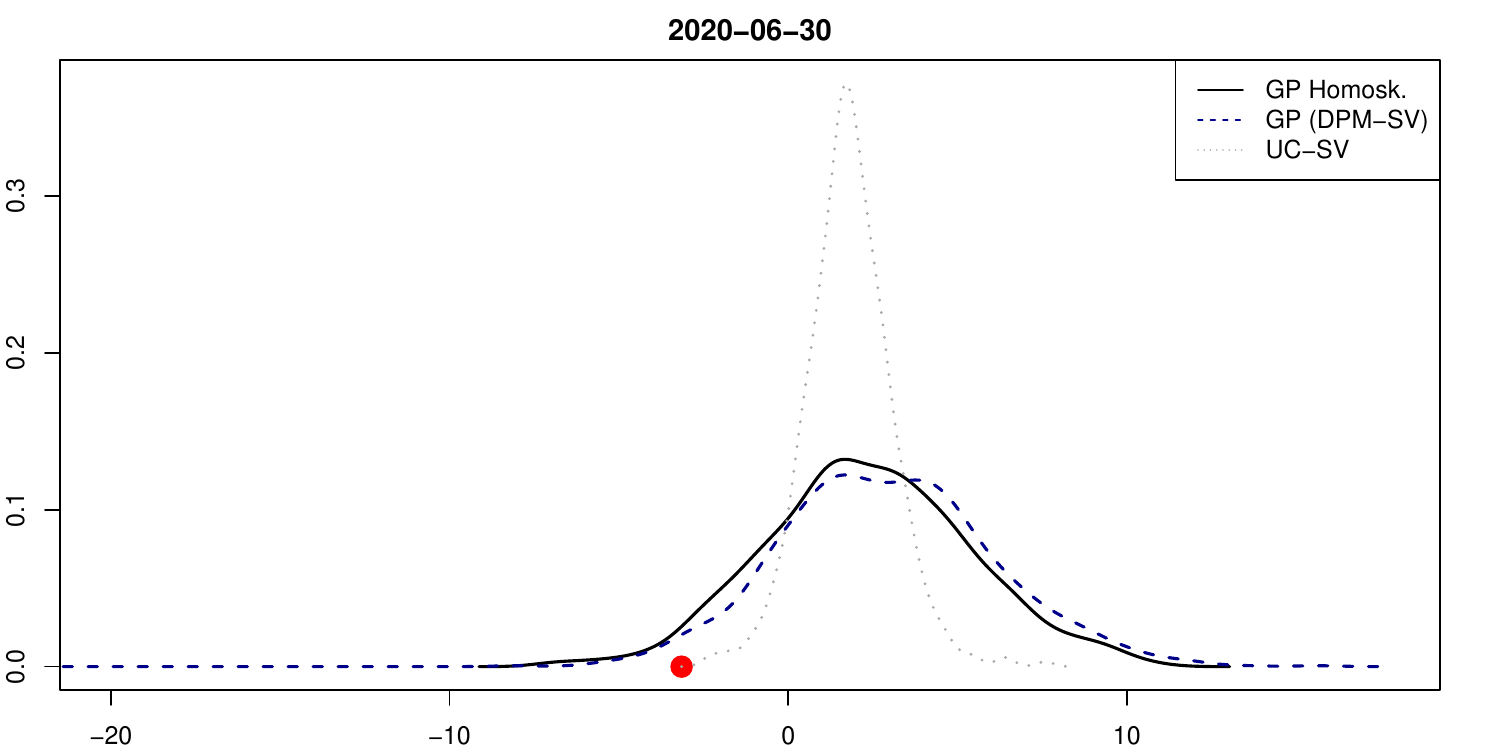}
    \includegraphics[page=157, scale=0.3]{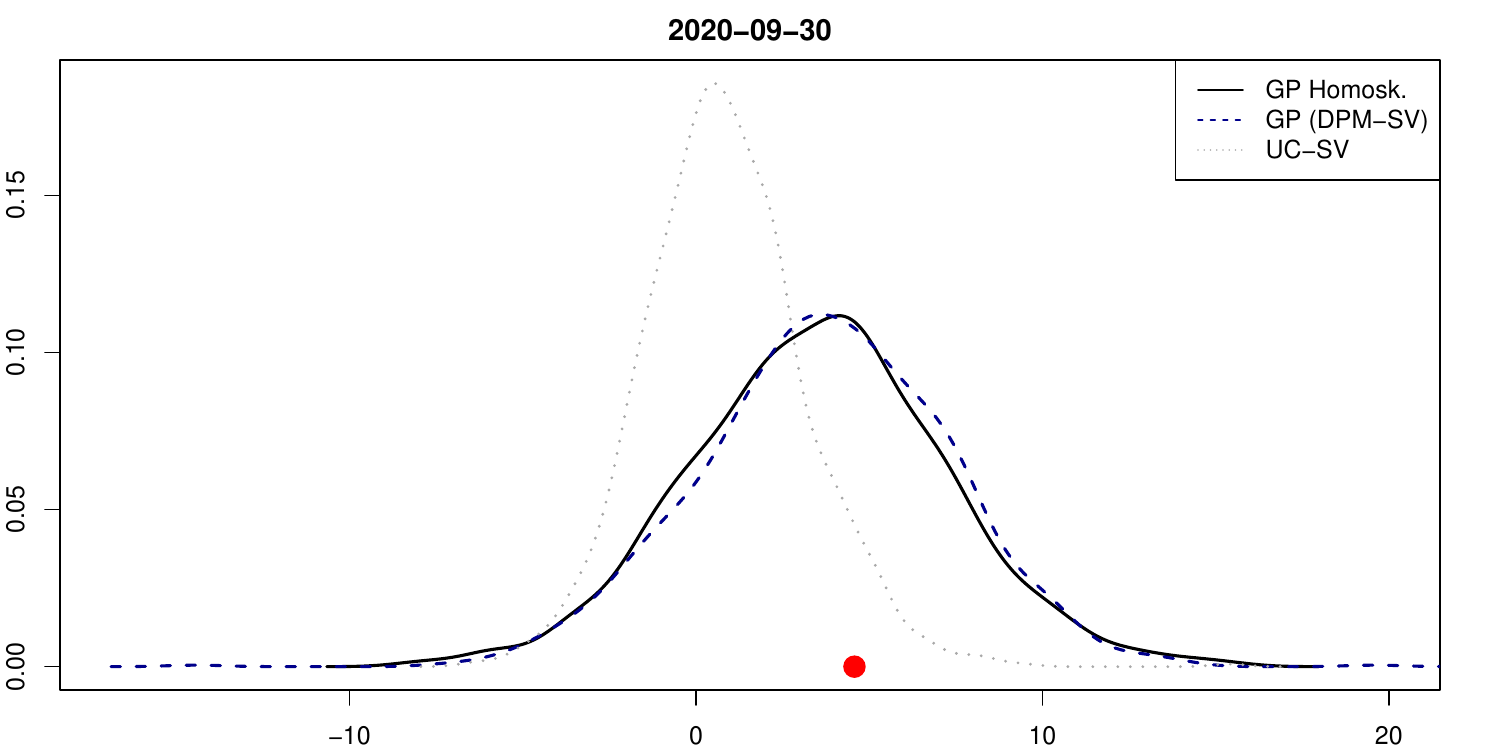}
    \includegraphics[page=158, scale=0.3]{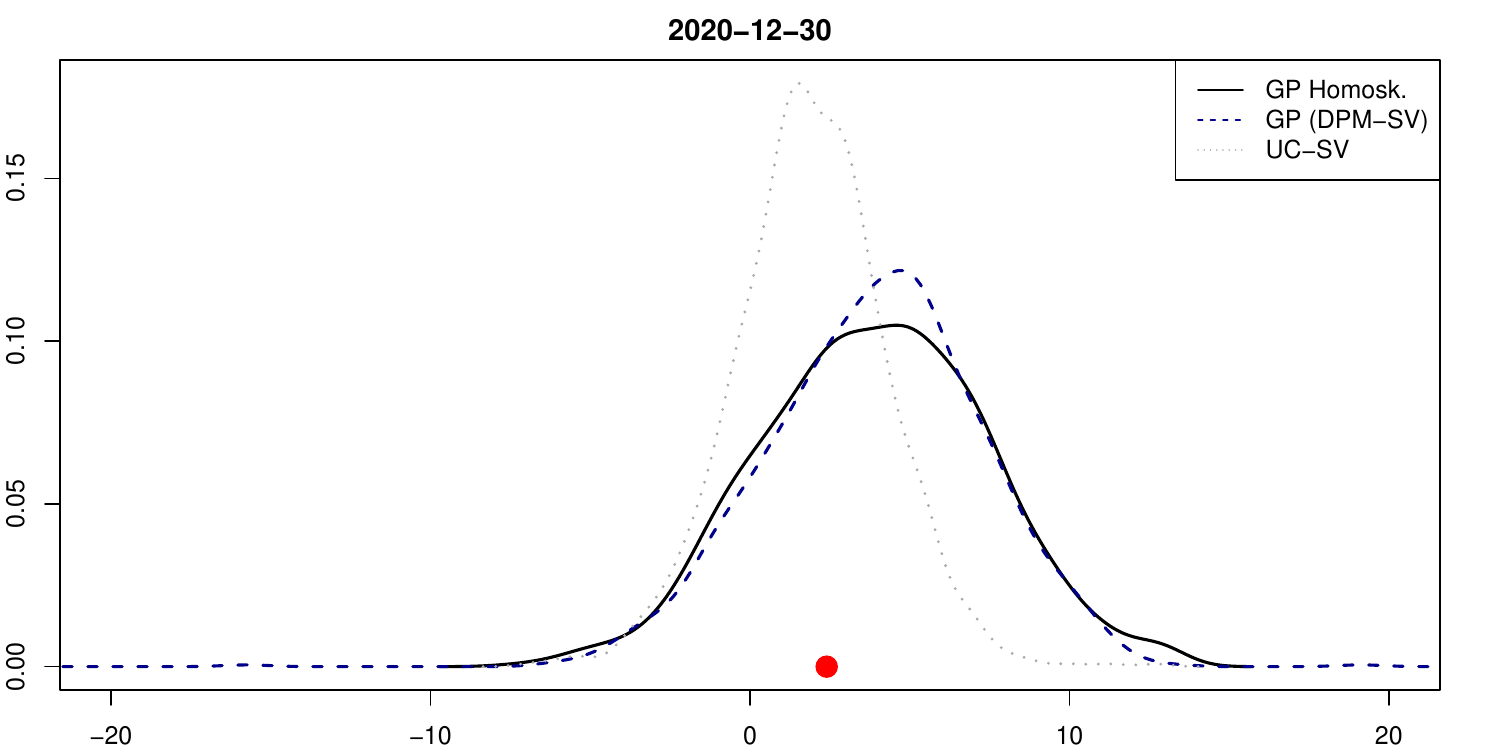}
    \end{minipage}
       \caption{Predictive densities of GP (with DPM-SV) and UC-SV with the moderately sized data set: 2008Q3 - 2009Q2 (left) and 2020Q1 - 2020Q4 (right)}
    \label{fig:dens_GFC}
\end{figure}

To get a better understanding of the effects of extreme events such as the financial crisis and the pandemic on the predictive distributions,  Figure \ref{fig:dens_GFC} plots the predictive densities in some quarters of these specific episodes.  The figure provides results from the UC-SV benchmark and the homoskedastic and DPM-SV versions of the moderately sized GP specification.  In each case, the red dot provides the actual outcome of inflation for the quarter in question.

These plots nicely illustrate that, in some quarters (especially at the beginning of the aforementioned periods), the gains from the more flexible model stem from non-Gaussian features (heavy tails and skewness). After one or two quarters, the GP models also seem to adjust the point forecasts (better than the UC-SV models) and this provides further predictive gains.  Consider late 2008 and early 2009.  Inflation in 2008:Q4 plummeted to roughly -9 percent, far outside the predictive distributions based on data through 2008:Q3.  In response, in the subsequent quarter, the predictive distributions widened for all three models, and the predictive density from both GP specifications correctly shifted to the left more so than did the predicted densities of the UC-SV model. As another example, shown in Figure \ref{fig:dens_GFC}'s results for the first year of the pandemic, in the second quarter of 2020 the predictive distributions of the GP specifications correctly moved well to the left to capture inflation's sharp fall, whereas the UC-SV's distribution adapted much less and implied that the inflation outcome was in the far left tail.

\subsection{Which variables determine the predictive distribution?}
In a linear regression model, the marginal effect of any predictor on inflation is simply its regression coefficient. In our nonparametric model, the nonlinear interaction between $\bm y_{t+h}$ and $\bm x_t$ means no such simple summary of the effect of a predictor on inflation exists. The effect can depend on the magnitude of any or all of the predictors and can vary over quantiles.  Several recent papers \citep[see, e.g.,][]{crawford2019variable, woody2021model} address this issue through linear posterior summaries that are close (according to some metric) to the actual posterior distribution. 

In this paper, we follow a similar approach to \cite{woody2021model}. Our aim is to approximate the quantiles of the predictive distribution $Q_{p, t+h}$ using a linear and possibly sparse regression model. For each $p$, we solve the following optimization problem:
\begin{equation}
    \bm \beta_p^* = \argmin_{\bm \beta_p} \sum_{t=t_0}^T \left(Q_{p, t+h} - \bm \beta'_p \bm x_t\right)^2 + \lambda \sum_{j=1}^K |\beta_{p,j}|, \label{eq: LASSO}
\end{equation}
where $t_0$ marks the beginning of the hold-out period,  $\bm \beta_p = (\beta_{p,1}, \dots, \beta_{p,K})'$ is a linear set of coefficients, and $\lambda \ge 0$ is a penalty term that controls the trade-off between model fit and parsimony. This is a LASSO problem, and we decide on $\lambda$ through cross-validation. Notice that for each quantile we search for a linear representation that minimizes the squared errors between the quantile forecast and the linear predictor. This closely resembles a standard quantile regression model and, if repeated for every $p$, provides us with a linearized version of the predictive distribution that is simple to interpret.

\begin{table}[ht!]
\centering
\begin{tabular}{rrrrrr}
  \toprule
  $p=$ & 0.05 & 0.1 & 0.5 & 0.9 & 0.95 \\ 
  \midrule
$h=1$ & 0.65 & 0.65 & 0.65 & 0.65 & 0.65 \\ 
  $h=4$ & 0.39 & 0.47 & 0.66 & 0.61 & 0.65 \\ 
   \bottomrule
\end{tabular}\\
\begin{minipage}[t]{0.4\textwidth}
\footnotesize \textbf{Notes}: The R$^2$ values are computed based on $\bm \beta_p$ obtained from solving the minimization problem in Equation \ref{eq: LASSO}.
\end{minipage}
\captionsetup{justification=centering}
\caption{Quantile and horizon-specific R$^2$ values}\label{tab_qslin}
\end{table}

Empirically, we focus on the GP model, with DPM-SV, for the moderately sized data set, one of the preferred specifications for predicting US inflation based on the previous analysis. Table \ref{tab_qslin} presents the R$^2$ of the corresponding quantile-specific linear regression. The table  shows that the fit of the linearized version of the predictive distribution is reasonably good --- at least for the purpose of capturing key covariates, although not necessarily for capturing the distributional complexities we have emphasized --- in particular for $h=1$. At this forecast horizon, we observe only little differences in R$^2$ across quantiles. When we focus on $h=4$, the explanatory power of the regression decreases (especially at the lower tails) but remains solid as an approximation. Specifically, while the R$^2$ is just below 0.4 for $p=0.05$, it increases to around $0.65$ for $p=0.95$. This indicates that when we focus on four quarters ahead, the linear model captures a significant portion of low-frequency movements in not only the left tail but also  the right tail.  In this respect, the linear model proves to be a useful approximation to our flexible nonparametric specification for the purpose of capturing key covariates.

\begin{figure}[t!]
    \centering
        \begin{minipage}[t]{0.46\textwidth}
        \hspace{2cm} \textbf{1-quarter-ahead}
\end{minipage}
        \begin{minipage}[t]{0.46\textwidth}
      \hspace{2cm}  \textbf{4-quarters-ahead}
\end{minipage}
    \includegraphics[trim=13 30 0 0,clip, scale=0.95]{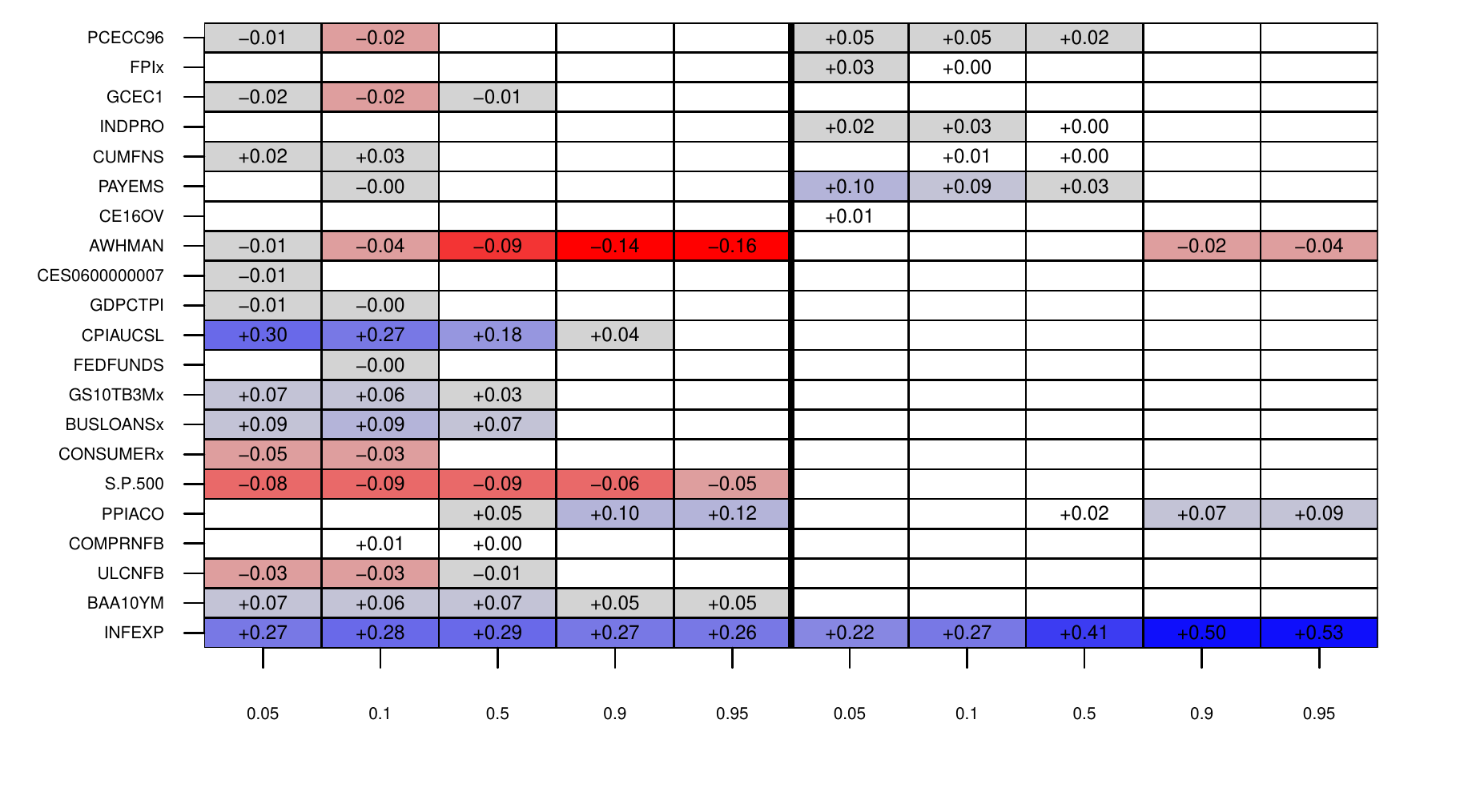}
    \begin{minipage}[t]{1\textwidth}
\footnotesize \textbf{Notes}: The heatmap shows standardized coefficients for quantiles $p \in \{0.05, 0.10, 0.5, 0.9, 0.95\}$. Different shadings of red indicate negative coefficients, whereas different shadings of blue indicate positive coefficients. Gray cells refer to small but non-zero coefficients. 
\end{minipage}

    \caption{Predictive variable relevance over quantiles: Moderately sized GP regression with DPM-SV}
    \label{fig:heatmap}
\end{figure}

Figure \ref{fig:heatmap} reports the estimated coefficients of the significant variables in the quantile-LASSO model (after standardizing all variables so that coefficients are comparable across regressors). Several interesting findings emerge. First, inflation expectations are consistently helpful, particularly so for $h=4$ and in the center and right tail of the distribution. Second, the autoregressive lag does not matter for $h=4$, and for $h=1$ it is only relevant in the middle and left tail of the distribution. Third, lagged inflation and inflation expectations are the variables with the largest coefficients, in line with the New Keynesian Phillips curve. The latter also suggests a role for real and labor market variables, and we find that for $h=1$, hours worked matter in the center and right tail of the distribution, but they appear with a negative sign.  However, employment has the correct sign for $h=4$ and is relevant for the left tail of the predictive distribution of inflation. Fourth, some financial indicators are also relevant, in particular corporate and term spreads for $h=1$ and in the left tail, where business loans also enter with a positive sign and the S\&P500 with a negative one. Fifth, PPI information is particularly useful in the right tail, for both $h=1$ and $h=4$. This can proxy for cost-push factors, such as increases in the prices of energy or imported goods. Finally, and in summary, for $h=1$, lagged and expected inflation are particularly relevant for the left tail of the predictive distribution of inflation, together with a set of financial indicators. In the right tail expected but not lagged inflation matters, together with PPI and hours worked. For $h=4$, fewer indicators are relevant: expectations and employment in the left tail, and expectations and PPI in the right tail. Overall, while our flexible nonparametric models capture the time-varying complexities of inflation's predictive distribution, the predictive distributions can be related to modest numbers of indicators commonly thought to have predictive content for inflation.

\section{Forecasting core inflation}\label{sec:coreinflation}

As a robustness check, we assess whether our model is also able to compete with the UC-SV benchmark when the focus is on core CPI (i.e., CPI excluding food and energy) inflation. This exercise is considerably harder for the models we propose (which are designed to capture non-systematic, higher-frequency movements in inflation), since core CPI inflation excludes the volatile food and energy components.
\begin{table}[h!]
\centering
\scalebox{0.75}{
\begin{tabular}{llllllllll}
 \hline
   &\multicolumn{4}{c}{Full hold-out period} & & \multicolumn{4}{c}{Only pandemic observations}\\
  &\multicolumn{4}{c}{(1980:Q1-2021:Q3)} & & \multicolumn{4}{c}{(2020:Q1 to 2021:Q3)}\\
  \midrule
 & Homosk. & DPM & SV & DPM-SV &  & Homosk. & DPM & SV & DPM-SV \\ 
    \midrule
     \multicolumn{10}{c}{Unobserved components models}\\ 
  $h=1$ & 1.467 & 0.985 & 0.417 & 0.986 &  & 1.147 & 0.95 & 1.077 & 0.965 \\ 
   & (-1.084) & (-0.392) & (-0.275) & (-0.076) &  & (-1.296) & (1.469) & (-4.868) & (0.62) \\ 
  $h=4$ & 1.471 & 1.065 & 0.396 & 1.001 &  & 1.934 & 1.289 & 0.395 & 1.54 \\ 
   & (-0.957) & (-0.819) & (-0.156) & (-0.187) &  & (0.82) & (1.42) & (-2.211) & (1.296) \\ 
        \midrule
     \multicolumn{10}{c}{AR(1) models}\\ 
  $h=1$ &  &  &  &  &  &  &  &  &  \\ 
  Linear & 2.194 & 1.844 & 1.321 & 1.439 &  & 0.96 & 0.928 & 0.89 & 0.929 \\ 
   & (-1.066) & (-0.758) & (-0.398) & (-0.499) &  & (3.424) & (3.399) & (2.998) & (3.107) \\ 
  GP & 2.285 & 1.937 & 1.405 & 1.515 &  & 0.92 & 0.913 & 0.917 & 0.887 \\ 
   & (-1.048) & (-0.809) & (-0.419) & (-0.55) &  & (3.482) & (3.34) & (3.062) & (3.165) \\ 
  GP-sub & 2.228 & 1.885 & 1.355 & 1.475 &  & 0.966 & 0.894 & 0.922 & 0.912 \\ 
   & (-1.05) & (-0.787) & (-0.383) & (-0.514) &  & (3.439) & (3.301) & (3.131) & (3.089) \\ 
  $h=4$ &  &  &  &  &  &  &  &  &  \\ 
  Linear & 2.366 & 2.003 & 1.339 & 1.448 &  & 1.914 & 1.385 & 0.711 & 0.732 \\ 
   & (-1.205) & (-0.914) & (-0.573) & (-0.648) &  & (1.013) & (1.455) & (1.538) & (1.68) \\ 
  GP & 2.47 & 2.139 & 1.382 & 1.534 &  & 2.127 & 1.6 & 0.689 & 0.942 \\ 
   & (-1.205) & (-0.976) & (-0.585) & (-0.669) &  & (0.974) & (1.349) & (1.934) & (1.958) \\ 
  GP-sub & 2.379 & 2.01 & 1.365 & 1.529 &  & 1.932 & 1.681 & 0.734 & 0.787 \\ 
   & (-1.182) & (-0.916) & (-0.566) & (-0.673) &  & (1.03) & (1.278) & (1.192) & (1.636) \\ 
           \midrule 
     \multicolumn{10}{c}{Moderately sized models}\\ 
  $h=1$ &  &  &  &  &  &  &  &  &  \\ 
  Linear & 1.168 & 1.228 & 1.362 & 1.321 &  & 0.827 & 0.845 & 0.913 & 0.924 \\ 
   & (-0.391) & (-0.448) & (-0.515) & (-0.44) &  & (3.108) & (2.842) & (0.854) & (1.677) \\ 
  GP & 1.043 & 1.074 & 0.997 & 1.009 &  & 0.954 & 0.98 & 0.939 & 0.974 \\ 
   & (-0.269) & (-0.322) & (0.059) & (-0.072) &  & (3.079) & (3.414) & (2.753) & (3.633) \\ 
  GP-sub & 1.012 & 1.038 & 1.079 & 1.036 &  & 0.908 & 0.934 & 0.877 & 0.915 \\ 
   & (-0.271) & (-0.335) & (-0.427) & (-0.064) &  & (2.9) & (2.883) & (-3.484) & (3.465) \\ 
  $h=4$ &  &  &  &  &  &  &  &  &  \\ 
  Linear & 1.196 & 1.385 & 1.044 & 1.044 &  & 1.274 & 1.4 & 0.977 & 1.266 \\ 
   & (-0.562) & (-0.711) & (-0.626) & (-0.435) &  & (1.492) & (1.341) & (-1.449) & (0.795) \\ 
  GP & 0.933 & 1.012 & 0.837 & 0.848 &  & 0.805 & 0.899 & 0.671 & 0.679 \\ 
   & (-0.3) & (-0.384) & (0.052) & (-0.148) &  & (1.914) & (1.869) & (2.161) & (2.045) \\ 
  GP-sub & 0.969 & 1.052 & 0.81 & 0.928 &  & 0.949 & 1.02 & 0.77 & 1.015 \\ 
   & (-0.313) & (-0.425) & (0.021) & (-0.107) &  & (1.902) & (1.792) & (2.166) & (2.013) \\ 
              \midrule
     \multicolumn{10}{c}{Large-scale models}\\ 
  $h=1$ &  &  &  &  &  &  &  &  &  \\ 
  Linear & 1.11 & 1.182 & 1.134 & 1.115 &  & 0.873 & 0.976 & 0.936 & 0.949 \\ 
   & (-0.45) & (-0.414) & (-0.208) & (-0.221) &  & (3.283) & (2.436) & (3.298) & (3.346) \\ 
  GP & 1.117 & 1.175 & 1.017 & 1.069 &  & 0.851 & 0.921 & 0.845 & 0.856 \\ 
   & (-0.366) & (-0.44) & (-0.12) & (-0.21) &  & (3.571) & (3.505) & (1.572) & (3.599) \\ 
  GP-sub & 1.053 & 1.201 & 0.976 & 1.026 &  & 0.845 & 0.897 & 0.856 & 0.86 \\ 
   & (-0.347) & (-0.443) & (-Inf) & (-0.123) &  & (3.373) & (3.479) & (-Inf) & (3.044) \\ 
  $h=4$ &  &  &  &  &  &  &  &  &  \\ 
  Linear & 0.999 & 1.241 & 1.042 & 1.039 &  & 0.744 & 1.147 & 0.884 & 0.97 \\ 
   & (-0.529) & (-0.502) & (-0.356) & (-0.344) &  & (1.727) & (1.817) & (1.822) & (1.874) \\ 
  GP & 1.201 & 1.251 & 1.012 & 1.07 &  & 0.909 & 0.972 & 0.62 & 0.661 \\ 
   & (-0.522) & (-0.578) & (-0.091) & (-0.321) &  & (1.675) & (1.656) & (1.896) & (1.889) \\ 
  GP-sub & 0.994 & 1.236 & 0.843 & 0.923 &  & 0.765 & 0.747 & 0.706 & 0.734 \\ 
   & (-0.451) & (-0.56) & (-0.04) & (-0.21) &  & (1.782) & (1.744) & (2.068) & (2.044) \\ 
   \hline
\end{tabular}
}
\caption{MSE and Average LPL Results for Core Inflation.  LPL results given in parentheses. Results are relative (MSE ratios or LPL differences) to the UC-SV benchmark.}\label{core}
\end{table}

Table \ref{core} shows the results for core CPI inflation.  In these results, as in the results for headline inflation, the MSE for the benchmark UC-SV model is much higher in the pandemic period than in the full sample at the one-step-ahead horizon, whereas at the four-steps-ahead horizon, the MSE is more stable across samples.  But for both horizons, the LPL for the benchmark model drops substantially in the pandemic period compared to the full sample.  In the full sample, using core rather than headline inflation reduces the advantages of the GP and GP-sub specifications compared to the UC-SV benchmark.   Among these models, the SP specification with SV applied to the moderately sized set of variables fares best over the full sample, about the same as the benchmark in MSE and LPL accuracy for $h=1$ and modestly to solidly more accurate for $h=4$.  Over the pandemic sample, most models beat the UC-SV benchmark in both MSE and LPL.  In particular, the GP models offer gains larger than those found over the long sample, with the SV specification again best with GP.  In this sample, by the MSE metric, gains are larger for $h=4$ (as large as 33 percent) than for $h=1$.  In addition, over the pandemic, there is some additional advantage to using the large-scale variable set as compared to the medium-scale data set.

\section{Conclusions}\label{sec:concl}
Forecasting inflation is hard, partly due to the changing, and often vanishing, relationship between it and its predictors and partly due to the occasionally large and asymmetric shocks that hit the inflation process. The model developed in this paper, being nonparametric in the conditional mean and in the error distribution, is designed to address these challenges. 

In our empirical work, we have shown that the model is capable of producing accurate point and density forecasts. These forecasts are often more precise than the ones obtained from simpler alternatives such as the UC-SV model or a linear regression model. This forecasting performance is driven by a superior  overall performance in the left tail and the center of the distribution. However, the performance in the right tail is somewhat weaker. This is mainly driven by slightly inflated predictive intervals during the Great Moderation. In the high-inflation period of the early 1980s and during the second year of the pandemic, our model also improves upon the UC-SV model in the right tail.

Given that the model works well in turbulent times, it might also work well for forecasting time series that are subject to rapid mean shifts and changing volatilities such as exchange rates or asset prices. Moreover, the univariate nature of the model implies that we do not model dynamic interdependencies between variables. Since our framework is highly scalable, it would be straightforward to extend it to the VAR case and use it for structural analysis. 

%We have found it to  produce forecasts which, in most dimensions, are superior to common alternatives. Of particular note is its ability to improve forecasts during the pandemic and in the left tail.

\newpage
\small{\setstretch{0.85}
\addcontentsline{toc}{section}{References}
\bibliographystyle{frbcle.bst}
\bibliography{lit}}\normalsize\clearpage

\begin{appendices}
\renewcommand{\thefigure}{\thesection \arabic{figure}}
\renewcommand{\thetable}{\thesection \arabic{table}}
\renewcommand{\theequation}{\thesection .\arabic{equation}}

\setcounter{table}{0}
\setcounter{equation}{0}
\setcounter{figure}{0}

\section{Full conditional posterior simulation}\label{app:MCMC}
In this section we discuss our posterior simulation algorithm. With some exceptions, the parameters and states of the model can be sampled through  Gibbs updating steps.  In cases where the full conditionals are not of a well-known form, we rely on Metropolis-Hastings updates to simulate from the target distribution.

\subsection{Posterior simulation of the Dirichlet process mixture}\label{app: dpm}
We start by discussing the posterior simulation steps of the DPM first. This algorithm closely follows the one discussed in \cite{FS_MW}. Notice that (\ref{DPM}) can be written as:
\begin{equation}
    \varepsilon_t|\delta_t = j \sim \mathcal{N}(\mu_j, \sigma_j^2), \label{eq: mixRep}
\end{equation}
with $\delta_t$ denoting a latent variable that assigns each observation to one of the components. $\delta_t$ is specified such that $\text{Prob}(\delta_t = j) = w_j$ and we let $\bm \delta = (\delta_1, \dots, \delta_T)'$ denote a $T-$dimensional classification vector. In terms of full-data vectors, the DPM can be written as follows:
\begin{equation*}
    \bm \varepsilon = \bm \mu + \bm \eta, \quad \bm \eta \sim \mathcal{N}(\bm 0_T, \bm \Sigma),
\end{equation*}
with $\bm \varepsilon=(\varepsilon_1, \dots, \varepsilon_T)', \bm \mu = (\mu_{\delta_1}, \dots, \mu_{\delta_T})$ and $\bm \Sigma = \text{diag}(\sigma^2_{\delta_1}, \dots,\sigma^2_{\delta_T})$. %This representation allows us to derive a Gibbs sampler. 

To simulate from the posterior of the DPM, we iterate between the following steps (conditional on knowing $\bm f$):
\begin{itemize}
    \item We start by sampling the weights $\xi_i$ conditional on the classification indicators and the remaining model parameters and latent states. Let $\xi_K =1$ (where $J$ is a truncation level for the number of mixture components). The sticks $\xi_1, \dots, \xi_{J-1}$ are simulated from a sequence of Beta distributions:
\begin{equation}
    \xi_j|\bm \delta \sim \mathcal{B}\left(1 + T_j, \alpha + \sum_{i=j+1}^J T_i\right),
\end{equation}
where $T_j = \sum_{t=1}^T \mathbb{I}(\delta_t=j)$ denotes the number of observations associated with cluster $j$, with $\mathbb{I}(\bullet)$ denoting an indicator function that equals $1$ if its argument is true. Draws of $\xi_1, \dots, \xi_{J-1}$ are then used to construct the weights $w_1, \dots, w_{J-1}$ using the stick-breaking formula in (\ref{eq: stickbreaks}).

\item Next we sample the classification indicators on a $t-$by$-t$ basis using slice sampling \citep{kalli2011slice}. We do this in two steps. First, we sample an auxiliary quantity $u_t|\delta_t \sim \mathcal{U}(0, \varpi_{\delta_t})$ with $\varpi_{j} = (1 - \kappa) \kappa^{j-1}$  and $\kappa =0.8$. In the second step, the indicators are simulated from a discrete distribution with the probability that $\delta_t=j~(j=1, \dots, J)$ being proportional to:
\begin{equation*}
   \text{Prob}(\delta_t=j|\bullet) \frac{\mathbb{I}(u_t < \varpi_j)}{\varpi_j} \times w_j \mathcal{N}(\epsilon_t|\mu_j, \sigma_j^2).
\end{equation*}
The $\bullet$ notation indicates conditioning on all remaining coefficients and latent states of the model.

\item The posterior of $\alpha$ is obtained through a  Metropolis-Hastings  step \citep[see][]{FS_MW}.

\item The component mean $\mu_j~(j=1, \dots, J)$ is simulated from:
\begin{equation*}
    \mu_j|\bullet \sim \mathcal{N}(\overline{\mu}_j, \overline{v}_j)
\end{equation*}
with 
\begin{align*}
    \overline{v}_j &= (T_j + \underline{v}^{-1}_j)^{-1}, \\
    \overline{\mu}_j &=  \overline{v}_j \sum_{t=1}^T \left(\varepsilon_t \times \mathbb{I}(\delta_t = j)\right).
\end{align*}

\item We simulate the component-specific error variances from an inverse Gamma-distributed posterior:
\begin{equation*}
    \sigma_j^2|\bullet \sim \mathcal{G}^{-1}\left(T_j + c_0, c_1 + \frac{\sum_{t=1}^T[(\varepsilon_t - \mu_{\delta_t})^2 \mathbb{I}(\delta_t=j)] }{2}\right).
\end{equation*}
\end{itemize}
Finally, we obtain the truncation level $J$ such that $1 - \sum_{j=1}^J w_j < \text{min}(u_1, \dots, u_T)$.

In case we use the stochastic volatility specification, the sampling step for the component-specific variances is replaced with the algorithm proposed in \cite{kastner2014ancillarity}. 

\subsection{Posterior simulation of the GP regression}\label{app: GP}
Conditional on $\bm \mu$ and $\bm \Sigma$, we need to sample $\bm f$ and the parameter controlling the weight put on the subspace shrinkage prior $\tau^2$:
\begin{itemize}
    \item The full conditional posterior of $\bm f$ is a $T$-dimensional Gaussian distribution:
\begin{equation*}
    \bm f | \bullet \sim \mathcal{N}(\overline{\bm f}, \overline{\bm V})
\end{equation*}
where
\begin{align*}
    \overline{\bm V} &=  \bm K_1 - \bm K_1 (\bm K_1 + \bm \Sigma)^{-1} \bm K'_1, \\
    \overline{\bm f} &= \bm K_1 (\bm K_1 + \bm \Sigma)^{-1} (\bm y - \bm \mu).
\end{align*}

\item We sample from the posterior of $\tau^2$ using a slice sampler \citep{subspace}. In the first step, we sample a uniformly distributed random variable $u \sim \mathcal{U}(0, r)$ with $r=(\tau^{-2} + 1)^{-(d_0 + d_1)^2}$ and set $r^* = u^{-(d_0+d_1)^{-1}}-1$. In the next step, we simulate yet another auxiliary quantity $\zeta$ from a truncated Gamma distribution with bounds $0$ and $r^*$:
\begin{equation*}
    \zeta|\bullet \sim \mathcal{G}_{0, r^*}\left(d_0 + \frac{T-K}{2}, \frac{{\bm f}' (\bm I - \bm \Psi_0) {\bm f}}{2} \right).
\end{equation*}
 A draw of $\tau^2$ is then obtained by setting $\tau^2 = 1/\zeta$.

\item We estimate $\phi$ and $\xi$ jointly using a random walk Metropolis-Hastings updating step. 
\end{itemize}

\newpage
\section{Data}\label{app:data}
\begin{table}[h!]
{\tiny
\begin{center}
\scalebox{0.8}{
\begin{tabular}{lllccc}
\toprule
\multicolumn{1}{l}{\ }&\multicolumn{1}{c}{\ Mnemonic}&\multicolumn{1}{c}{\ Description}&\multicolumn{1}{c}{\ Trans}&\multicolumn{1}{c}{\ M}&\multicolumn{1}{c}{\ L}\tabularnewline
\midrule
{\scshape }&&&&&\tabularnewline
~~&INFEXP& 5-quarter-ahead inflation expectations from the Survey of Professional Forecasters&$1$&x&x\tabularnewline
~~&GDPC1&Real Gross Domestic Product&$5$&x&x\tabularnewline
~~&PCECC96&Real Personal Consumption Expenditures&$5$&x&x\tabularnewline
~~&FPIx&Real private fixed investment &$5$&x&x\tabularnewline
~~&GCEC1&Real Government Consumption Expenditures and Gross Investment&$5$&x&x\tabularnewline
~~&INDPRO&IP:Total index Industrial Production Index (Index 2012=100)&$5$&x&x\tabularnewline
~~&CUMFNS&Capacity Utilization:  Manufacturing (SIC) (Percent of Capacity)&$1$&x&x\tabularnewline
~~&PAYEMS& Emp:Nonfarm All Employees: Total nonfarm (Thousands of Persons)&$5$&x&x\tabularnewline
~~&CE16OV&Civilian Employment (Thousands of Persons)&$5$&x&x\tabularnewline
~~&UNRATE&Civilian Unemployment Rate (Percent)&$2$&x&x\tabularnewline
~~&AWHMAN&Average Weekly Hours of Production and Nonsupervisory Employees:  Manufacturing (Hours)&$1$&x&x\tabularnewline
~~&CES0600000007&Average Weekly Hours of Production and Nonsupervisory Employees:  Goods-Producing&$2$&x&x\tabularnewline
~~&CLAIMSx&Initial Claims&$5$&x&x\tabularnewline
~~&GDPCTPI&Gross Domestic Product: Chain-type Price Index&$6$&x&x\tabularnewline
~~&CPIAUCSL&Consumer Price Index for All Urban Consumers:  All Items&$6$&x&x\tabularnewline
~~&PPIACO&Producer Price Index for All Commodities &$6$&x&x\tabularnewline
~~&WPSID61&Producer Price Index by Commodity Intermediate Materials:  Supplies \& Components&$6$&x&x\tabularnewline
~~&WPSID62&Producer Price Index:  Crude Materials for Further Processing &$6$&x&x\tabularnewline
~~&COMPRNFB&Nonfarm Business Sector:  Real Compensation Per Hour (Index 2012=100)&$5$&x&x\tabularnewline
~~&ULCNFB&Nonfarm Business Sector:  Unit Labor Cost (Index 2012=100)&$5$&x&x\tabularnewline
~~&CES0600000008&Average Hourly Earnings of Production and Nonsupervisory Employees:&$6$&x&x\tabularnewline
~~&FEDFUNDS&Effective Federal Funds Rate (Percent)&$2$&x&x\tabularnewline
~~&BAA10YM&Moody's Seasoned Baa Corporate Bond Yield Relative to Yield on 10-Year Treasury&$1$&x&x\tabularnewline
~~&GS10TB3Mx&10-Year Treasury Constant Maturity Minus 3-Month Treasury Bill, secondary market&$1$&x&x\tabularnewline
~~&CPF3MTB3Mx&3-Month Commercial Paper Minus 3-Month Treasury Bill, secondary market&$1$&x&x\tabularnewline
~~&M2REAL&Real M2 Money Stock&$5$&x&x\tabularnewline
~~&BUSLOANSx&Real Commercial and Industrial Loans, All Commercial Banks&$5$&x&x\tabularnewline
~~&CONSUMERx&Real Consumer Loans at All Commercial Banks &$5$&x&x\tabularnewline
~~&S.P.500&S\& P's Common Stock Price Index:  Composite&$5$&x&x\tabularnewline
\midrule
{\scshape }&&&&&\tabularnewline
~~&PCDGx&Real personal consumption expenditures:  Durable goods &$5$&&x\tabularnewline
~~&PCESVx&Real Personal Consumption Expenditures:  Services &$5$&&x\tabularnewline
~~&PCNDx&Real Personal Consumption Expenditures:  Nondurable Goods &$5$&&x\tabularnewline
~~&GPDIC1&Real Gross Private Domestic Investment&$5$&&x\tabularnewline
~~&Y033RC1Q027SBEAx&Real Gross Private Domestic Investment:  Fixed Investment:  Nonresidential Equipment&$5$&&x\tabularnewline
~~&PNFIx&Real private fixed investment:  Nonresidential &$5$&&x\tabularnewline
~~&PRFIx&Real private fixed investment:  Residential &$5$&&x\tabularnewline
~~&A014RE1Q156NBEA&Shares of gross domestic product:  Gross private domestic investment: Change
in private inventories&$1$&&x\tabularnewline
~~&A823RL1Q225SBEA&Real Government Consumption Expenditures and Gross Investment:  Federal&$1$&&x\tabularnewline
~~&FGRECPTx&Real Federal Government Current Receipts &$5$&&x\tabularnewline
~~&SLCEx&Real government state and local consumption expenditures &$5$&&x\tabularnewline
~~&EXPGSC1&Real Exports of Goods and Services&$5$&&x\tabularnewline
~~&IMPGSC1&Real Imports of Goods and Services&$5$&&x\tabularnewline
~~&DPIC96&Real Disposable Personal Income&$5$&&x\tabularnewline
~~&OUTNFB&Nonfarm Business Sector:  Real Output&$5$&&x\tabularnewline
~~&OUTBS&Business Sector:  Real Output&$5$&&x\tabularnewline
~~&IPFINAL&IP:Final products Industrial Production: Final Products (Market Group) (Index 2012=100)&$5$&&x\tabularnewline
~~&IPCONGD&IP:Consumer goods Industrial Production: Consumer Goods (Index 2012=100)&$5$&&x\tabularnewline
~~&IPMAT&Materials (Index 2012=100)&$5$&&x\tabularnewline
~~&IPDMAT&Durable Materials (Index 2012=100)&$5$&&x\tabularnewline
~~&IPNMAT&Nondurable Materials (Index 2012=100)&$5$&&x\tabularnewline
~~&IPDCONGD&Durable Consumer Goods (Index 2012=100)&$5$&&x\tabularnewline
~~&IPB51110SQ&Durable Goods:  Automotive products (Index 2012=100)&$5$&&x\tabularnewline
~~&IPNCONGD&Nondurable Consumer Goods (Index 2012=100)&$5$&&x\tabularnewline
~~&IPBUSEQ&Business Equipment (Index 2012=100)&$5$&&x\tabularnewline
~~&IPB51220SQ&Consumer energy products (Index 2012=100)&$5$&&x\tabularnewline
~~&IPMANSICS&Industrial Production:  Manufacturing (SIC) (Index 2012=100)&$5$&&x\tabularnewline
~~&IPB51222S&Industrial Production:  Residential Utilities (Index 2012=100)&$5$&&x\tabularnewline
~~&IPFUELS&Industrial Production:  Fuels (Index 2012=100)&$5$&&x\tabularnewline
~~&USPRIV& All Employees: Total Private Industries (Thousands of Persons)&$5$&&x\tabularnewline
~~&MANEMP& All Employees: Manufacturing (Thousands of Persons)&$5$&&x\tabularnewline
~~&SRVPRD&All Employees:  Service-Providing Industries (Thousands of Persons)&$5$&&x\tabularnewline
~~&USGOOD&All Employees:  Goods-Producing Industries (Thousands of Persons)&$5$&&x\tabularnewline
~~&DMANEMP&All Employees:  Durable goods (Thousands of Persons)&$5$&&x\tabularnewline
~~&NDMANEMP&All Employees:  Nondurable goods (Thousands of Persons)&$5$&&x\tabularnewline
~~&USCONS&All Employees:  Construction (Thousands of Persons)&$5$&&x\tabularnewline
~~&USEHS&All Employees:  Education \& Health Services (Thousands of Persons)&$5$&&x\tabularnewline
~~&USFIRE&All Employees:  Financial Activities (Thousands of Persons)&$5$&&x\tabularnewline
~~&USINFO&All Employees:  Information Services (Thousands of Persons)&$5$&&x\tabularnewline
~~&USPBS&All Employees:  Professional \& Business Services (Thousands of Persons)&$5$&&x\tabularnewline
~~&USLAH&All Employees:  Leisure \& Hospitality (Thousands of Persons)&$5$&&x\tabularnewline
~~&USSERV&All Employees:  Other Services (Thousands of Persons)&$5$&&x\tabularnewline
~~&USMINE&All Employees:  Mining and logging (Thousands of Persons)&$5$&&x\tabularnewline
~~&USTPU&All Employees:  Trade, Transportation \& Utilities (Thousands of Persons)&$5$&&x\tabularnewline
~~&USGOVT&All Employees:  Government (Thousands of Persons)&$5$&&x\tabularnewline
~~&USTRADE&All Employees:  Retail Trade (Thousands of Persons)&$5$&&x\tabularnewline
~~&USWTRADE&All Employees:  Wholesale Trade (Thousands of Persons)&$5$&&x\tabularnewline
~~&CES9091000001&All Employees:  Government:  Federal (Thousands of Persons)&$5$&&x\tabularnewline
~~&CES9092000001&All Employees:  Government:  State Government (Thousands of Persons)&$5$&&x\tabularnewline
~~&CES9093000001&All Employees:  Government:  Local Government (Thousands of Persons)&$5$&&x\tabularnewline
~~&CIVPART&Civilian Labor Force Participation Rate (Percent)&$2$&&x\tabularnewline
~~&UNRATESTx&Unemployment Rate less than 27 weeks (Percent)&$2$&&x\tabularnewline
~~&UNRATELTx&Unemployment Rate for more than 27 weeks (Percent)&$2$&&x\tabularnewline
~~&LNS14000012&Unemployment Rate - 16 to 19 years (Percent)&$2$&&x\tabularnewline
~~&LNS14000025&Unemployment Rate - 20 years and over, Men (Percent)&$2$&&x\tabularnewline
~~&LNS14000026&Unemployment Rate - 20 years and over, Women (Percent)&$2$&&x\tabularnewline
\bottomrule
\end{tabular}}
%\caption{Data description.\label{tab:data1}}
\end{center}}
\end{table}
\begin{table}[h!]
{\tiny
\begin{center}
\scalebox{0.8}{
\begin{tabular}{lllccc}
\toprule
\multicolumn{1}{l}{\ }&\multicolumn{1}{c}{\ FRED.Mnemonic}&\multicolumn{1}{c}{\ Description}&\multicolumn{1}{c}{\ Trans}&\multicolumn{1}{c}{\ M}&\multicolumn{1}{c}{\ L}\tabularnewline
\midrule
{\scshape }&&&&&\tabularnewline
~~&UEMPLT5&Number of Civilians Unemployed - Less Than 5 Weeks (Thousands of Persons)&$5$&&x\tabularnewline
~~&UEMP5TO14&Number of Civilians Unemployed for 5 to 14 Weeks (Thousands of Persons)&$5$&&x\tabularnewline
~~&UEMP15T26&Number of Civilians Unemployed for 15 to 26 Weeks (Thousands of Persons)&$5$&&x\tabularnewline
~~&UEMP27OV&Number of Civilians Unemployed for 27 Weeks and Over (Thousands of Persons)&$5$&&x\tabularnewline
~~&AWOTMAN&Average Weekly Overtime Hours of Production and Nonsupervisory Employees: Manufacturing (Hours)&$2$&&x\tabularnewline
~~&HWIx&Help-Wanted Index&$1$&&x\tabularnewline
~~&HOUST&Housing Starts: Total: New Privately Owned Housing Units Started&$5$&&x\tabularnewline
~~&HOUST5F&Privately Owned Housing Starts: 5-Unit Structures or More&$5$&&x\tabularnewline
~~&PERMIT&New Private Housing Units Authorized by Building Permits&$5$&&x\tabularnewline
~~&HOUSTMW&Housing Starts in Midwest Census Region (Thousands of Units)&$5$&&x\tabularnewline
~~&HOUSTNE&Housing Starts in Northeast Census Region (Thousands of Units)&$5$&&x\tabularnewline
~~&HOUSTS&Housing Starts in South Census Region (Thousands of Units)&$5$&&x\tabularnewline
~~&HOUSTW&Housing Starts in West Census Region (Thousands of Units)&$5$&&x\tabularnewline
~~&RSAFSx&Real Retail and Food Services Sales (Millions of Chained 2012 Dollars)&$5$&&x\tabularnewline
~~&AMDMNOx&Real Manufacturers' New Orders:  Durable Goods (Millions of 2012 Dollars)&$5$&&x\tabularnewline
~~&AMDMUOx&Real Value of Manufacturers' Unfilled Orders for Durable Goods Industries&$5$&&x\tabularnewline
~~&BUSINVx&Total Business Inventories (Millions of Dollars)&$5$&&x\tabularnewline
~~&ISRATIOx&Total Business:  Inventories to Sales Ratio&$2$&&x\tabularnewline
~~&PCECTPI&Personal Consumption Expenditures: Chain-type Price Index &$6$&&x\tabularnewline
~~&PCEPILFE&Personal Consumption Expenditures Excluding Food and Energy&$6$&&x\tabularnewline
~~&GPDICTPI&Gross Private Domestic Investment: Chain-type Price Index &$6$&&x\tabularnewline
~~&IPDBS&Business Sector:  Implicit Price Deflator (Index 2012=100)&$6$&&x\tabularnewline
~~&DGDSRG3Q086SBEA&Personal consumption expenditures:  Goods &$6$&&x\tabularnewline
~~&DDURRG3Q086SBEA&Personal consumption expenditures:  Durable goods &$6$&&x\tabularnewline
~~&DSERRG3Q086SBEA&Personal consumption expenditures:  Services &$6$&&x\tabularnewline
~~&DNDGRG3Q086SBEA&Personal consumption expenditures:  Nondurable goods&$6$&&x\tabularnewline
~~&DHCERG3Q086SBEA&Personal consumption expenditures:  Services:  Household consumption expenditures&$6$&&x\tabularnewline
~~&DMOTRG3Q086SBEA&Personal consumption expenditures:  Durable goods:  Motor vehicles and parts&$6$&&x\tabularnewline
~~&DFDHRG3Q086SBEA&Personal consumption expenditures:  Durable goods:  Furnishings and durable household equipment&$6$&&x\tabularnewline
~~&DREQRG3Q086SBEA&Personal consumption expenditures:  Durable goods:  Recreational goods and vehicles&$6$&&x\tabularnewline
~~&DODGRG3Q086SBEA&Personal consumption expenditures:  Durable goods:  Other durable goods&$6$&&x\tabularnewline
~~&DFXARG3Q086SBEA&Personal consumption expenditures:  Nondurable goods:  Food and beverages purchased for off-premises consumption&$6$&&x\tabularnewline
~~&DCLORG3Q086SBEA&Personal consumption expenditures:  Nondurable goods:  Clothing and footwear&$6$&&x\tabularnewline
~~&DGOERG3Q086SBEA&Personal consumption expenditures:  Nondurable goods:  Gasoline and other energy goods&$6$&&x\tabularnewline
~~&DONGRG3Q086SBEA&Personal consumption expenditures:  Nondurable goods:  Other nondurable goods&$6$&&x\tabularnewline
~~&DHUTRG3Q086SBEA&Personal consumption expenditures:  Services:  Housing and utilities&$6$&&x\tabularnewline
~~&DHLCRG3Q086SBEA&Personal consumption expenditures:  Services:  Health care&$6$&&x\tabularnewline
~~&DTRSRG3Q086SBEA&Personal consumption expenditures:  Transportation services&$6$&&x\tabularnewline
~~&DRCARG3Q086SBEA&Personal consumption expenditures: Recreation services&$6$&&x\tabularnewline
~~&DFSARG3Q086SBEA&Personal consumption expenditures:  Services:  Food services and accomodations&$6$&&x\tabularnewline
~~&DIFSRG3Q086SBEA&Personal consumption expenditures:  Financial services and insurance&$6$&&x\tabularnewline
~~&DOTSRG3Q086SBEA&Personal consumption expenditures:  Other services &$6$&&x\tabularnewline
~~&CPILFESL&Consumer Price Index for All Urban Consumers:  All Items Less Food \& Energy&$6$&&x\tabularnewline
~~&WPSFD49207&Producer Price Index by Commodity for Finished Goods &$6$&&x\tabularnewline
~~&WPSFD49502&Producer Price Index by Commodity for Finished Consumer Goods &$6$&&x\tabularnewline
~~&WPSFD4111&Producer Price Index by Commodity for Finished Consumer Foods&$6$&&x\tabularnewline
~~&PPIIDC&Producer Price Index by Commodity Industrial Commodities &$6$&&x\tabularnewline
~~&WPU0561&Producer Price Index by Commodity for Fuels and Related Products and Power&$5$&&x\tabularnewline
~~&OILPRICEx&Real Crude Oil Prices:  West Texas Intermediate (WTI) - Cushing, Oklahoma&$5$&&x\tabularnewline
~~&PPICMM&Producer Price Index:  Commodities:  Metals and metal products:  Primary nonferrous metals&$6$&&x\tabularnewline
~~&CPIAPPSL&Consumer Price Index for All Urban Consumers:  Apparel&$6$&&x\tabularnewline
~~&CPITRNSL&Consumer Price Index for All Urban Consumers:  Transportation&$6$&&x\tabularnewline
~~&CPIMEDSL&Consumer Price Index for All Urban Consumers:  Medical Care&$6$&&x\tabularnewline
~~&CUSR0000SAC&Consumer Price Index for All Urban Consumers:  Commodities&$6$&&x\tabularnewline
~~&CES2000000008x&Real Average Hourly Earnings of Production and Nonsupervisory Employees: Construction&$5$&&x\tabularnewline
~~&CES3000000008x&Real Average Hourly Earnings of Production and Nonsupervisory Employees: Manufacturing&$5$&&x\tabularnewline
~~&TB3MS&3-Month Treasury Bill: Secondary Market Rate (Percent)&$2$&&x\tabularnewline
~~&TB6MS&6-Month Treasury Bill: Secondary Market Rate (Percent)&$2$&&x\tabularnewline
~~&GS1&1-Year Treasury Constant Maturity Rate (Percent)&$2$&&x\tabularnewline
~~&GS10&10-Year Treasury Constant Maturity Rate (Percent)&$2$&&x\tabularnewline
~~&AAA&Moody's Seasoned Aaa Corporate Bond Yield (Percent)&$2$&&x\tabularnewline
~~&BAA&Moody's Seasoned Baa Corporate Bond Yield (Percent)&$2$&&x\tabularnewline
~~&TB6M3Mx&6-Month Treasury Bill Minus 3-Month Treasury Bill, secondary market (Percent)&$1$&&x\tabularnewline
~~&GS1TB3Mx&1-Year Treasury Constant Maturity Minus 3-Month Treasury Bill, secondary market&$1$&&x\tabularnewline
~~&GS5&5-Year Treasury Constant Maturity Rate&$2$&&x\tabularnewline
~~&TB3SMFFM&3-Month Treasury Constant Maturity Minus Federal Funds Rate&$1$&&x\tabularnewline
~~&T5YFFM&5-Year Treasury Constant Maturity Minus Federal Funds Rate&$1$&&x\tabularnewline
~~&AAAFFM&Moody's Seasoned Aaa Corporate Bond Minus Federal Funds Rate&$1$&&x\tabularnewline
~~&M1REAL& Real M1 Money Stock&$5$&&x\tabularnewline
~~&NONREVSLx&Total Real Nonrevolving Credit Owned and Securitized, Outstanding&$5$&&x\tabularnewline
~~&REALLNx&Real Real Estate Loans, All Commercial Banks&$5$&&x\tabularnewline
~~&TOTALSLx&Total Consumer Credit Outstanding&$5$&&x\tabularnewline
~~&TOTRESNS&Total Reserves of Depository Institutions &$6$&&x\tabularnewline
~~&NONBORRES&Reserves Of Depository Institutions, Nonborrowed&$7$&&x\tabularnewline
~~&DTCOLNVHFNM&Consumer Motor Vehicle Loans Outstanding Owned by Finance Companies&$6$&&x\tabularnewline
~~&DTCTHFNM&Total Consumer Loans and Leases Outstanding Owned and Securitized by Finance Companies &$6$&&x\tabularnewline
~~&INVEST&Securities in Bank Credit at All Commercial Banks &$6$&&x\tabularnewline
~~&TABSHNOx&Real Total Assets of Households and Nonprofit Organizations&$5$&&x\tabularnewline
~~&EXSZUSx&Switzerland / U.S. Foreign Exchange Rate&$5$&&x\tabularnewline
~~&EXJPUSx&Japan / U.S. Foreign Exchange Rate&$5$&&x\tabularnewline
~~&EXUSUKx&U.S. / U.K. Foreign Exchange Rate&$5$&&x\tabularnewline
~~&EXCAUSx&Canada / U.S. Foreign Exchange Rate&$5$&&x\tabularnewline
~~&S.P..indust&S\& P's Common Stock Price Index:  Industrials&$5$&&x\tabularnewline
~~&S.P.div.yield&S\& P's Composite Common Stock:  Dividend Yield&$2$&&x\tabularnewline
\bottomrule
\end{tabular}}
\caption{Data overview. The column 'Trans' refers to the transformation codes according to the ones discussed in \cite{McCrackenNgFREDQD}. An 'X' in the column labeled M or L marks inclusion of a variable in the moderate or large data set.\label{tab:data_overview}}\end{center}}
\end{table}

\newpage
\section{Additional Empirical Results}
\setcounter{figure}{0}
\begin{figure}[t!]
    \centering
    \begin{minipage}[c]{1\linewidth}
       \textbf{(a) 1-quarter-ahead}
    \end{minipage}
    \begin{minipage}[c]{1\linewidth}
           \includegraphics[scale=0.51]{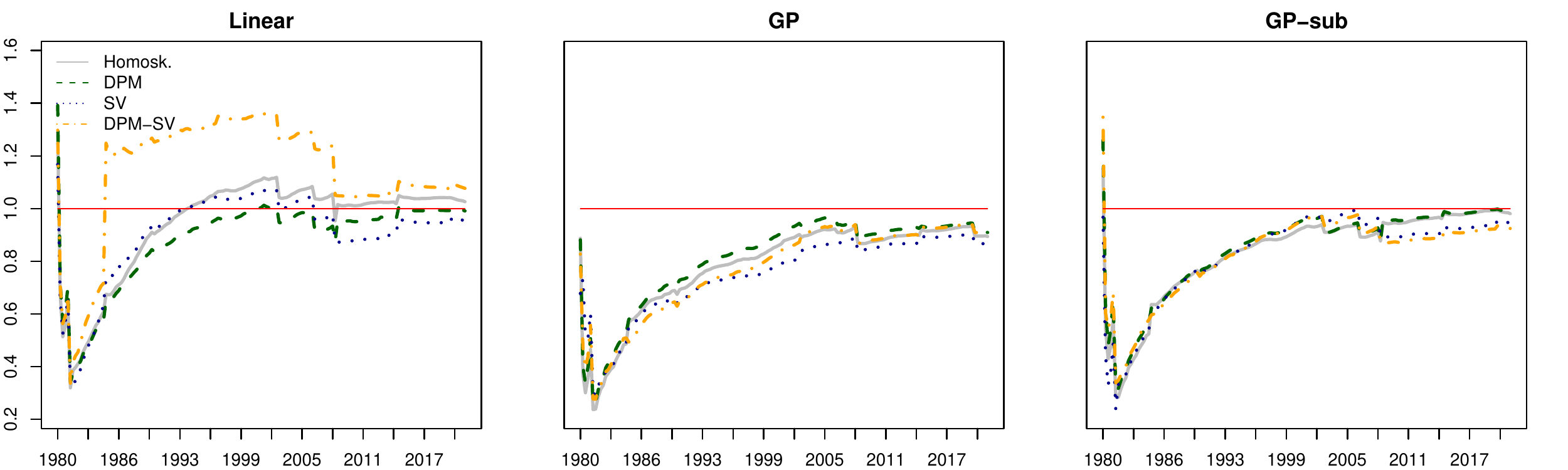}
    \end{minipage}
        \begin{minipage}[c]{1\linewidth}
        \vspace{0.2cm}
       \textbf{(a) 4-quarters-ahead}
    \end{minipage}
    \begin{minipage}[c]{1\linewidth}
           \includegraphics[scale=0.51]{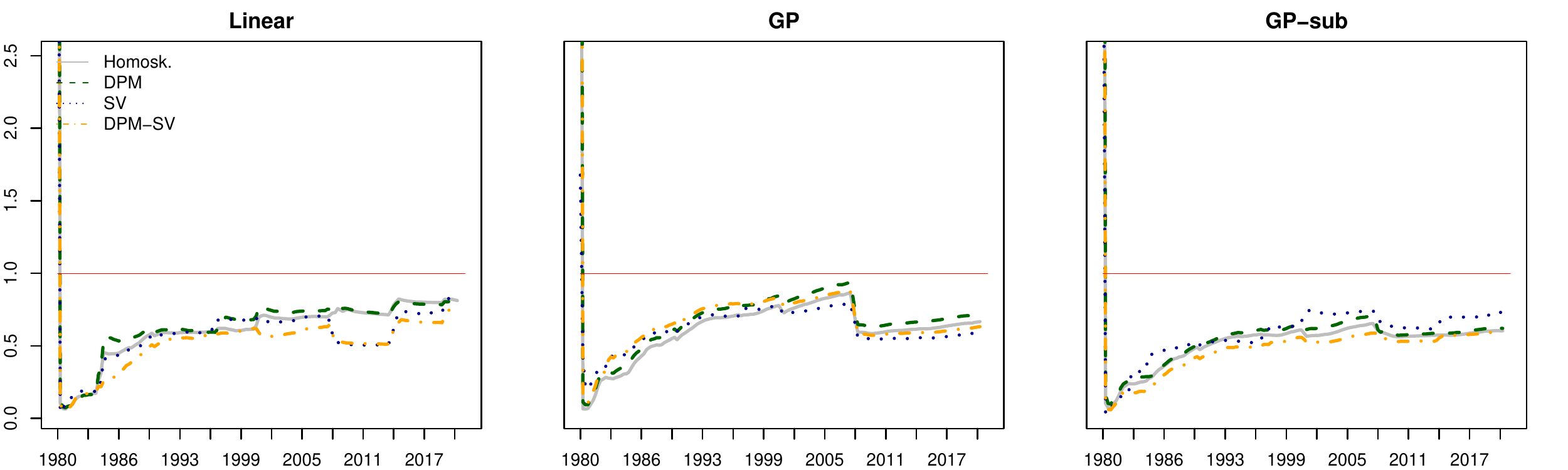}
    \end{minipage}
    \caption{Cumulative QS(0.05) against the UC-SV model: Moderately sized models}
    \label{fig:cumQS_small}
\end{figure}
\begin{figure}[h!]
    \centering
    \begin{minipage}[c]{1\linewidth}
       \textbf{(a) 1-quarter-ahead}
    \end{minipage}
    \begin{minipage}[c]{1\linewidth}
           \includegraphics[scale=0.51]{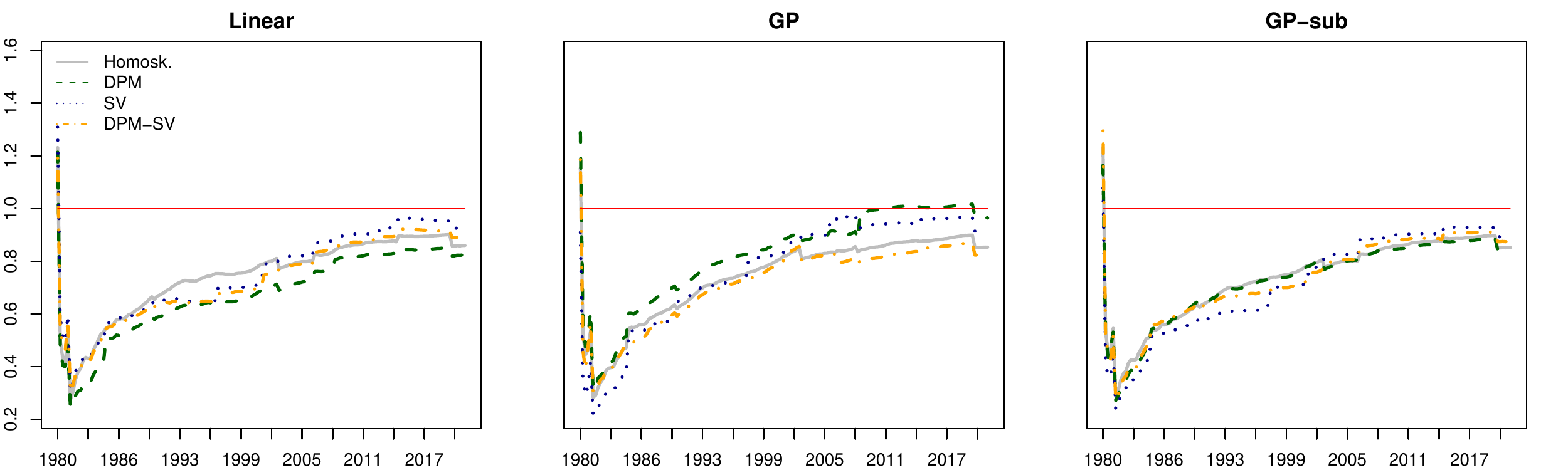}
    \end{minipage}
        \begin{minipage}[c]{1\linewidth}
        \vspace{0.2cm}
       \textbf{(a) 4-quarters-ahead}
    \end{minipage}
    \begin{minipage}[c]{1\linewidth}
           \includegraphics[scale=0.51]{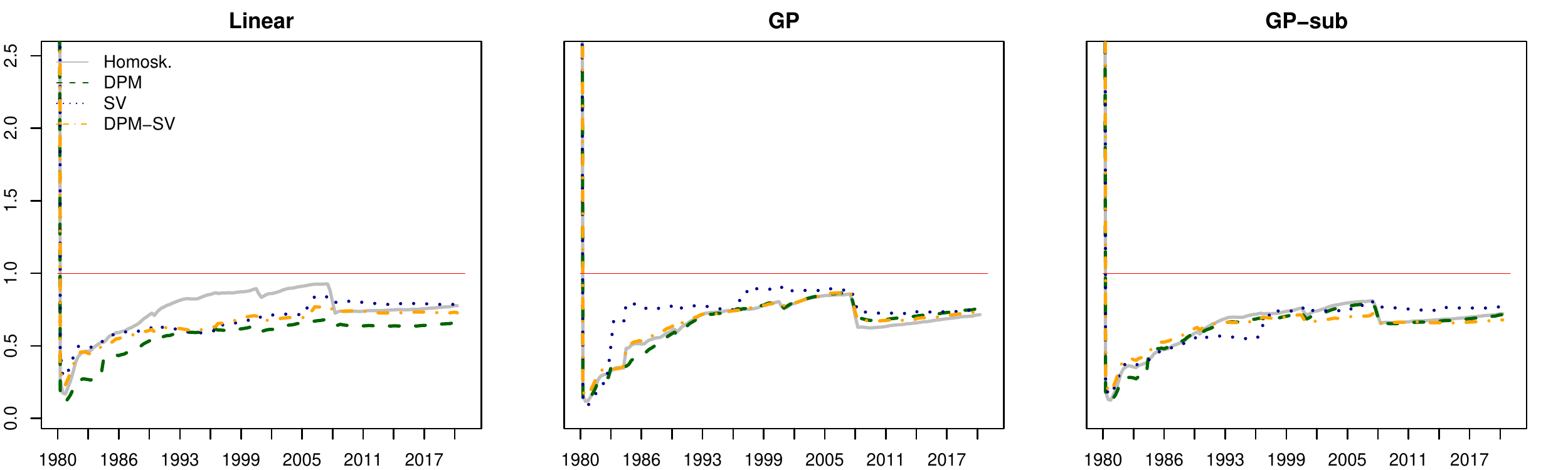}
    \end{minipage}
    \caption{Cumulative QS (0.05) against the UC-SV model: Large models}
    \label{fig:cumQS_small}
\end{figure}
\begin{figure}[h]
    \centering
    \begin{minipage}[c]{1\linewidth}
       \textbf{(a) 1-quarter-ahead}
    \end{minipage}
    \begin{minipage}[c]{1\linewidth}
           \includegraphics[scale=0.51]{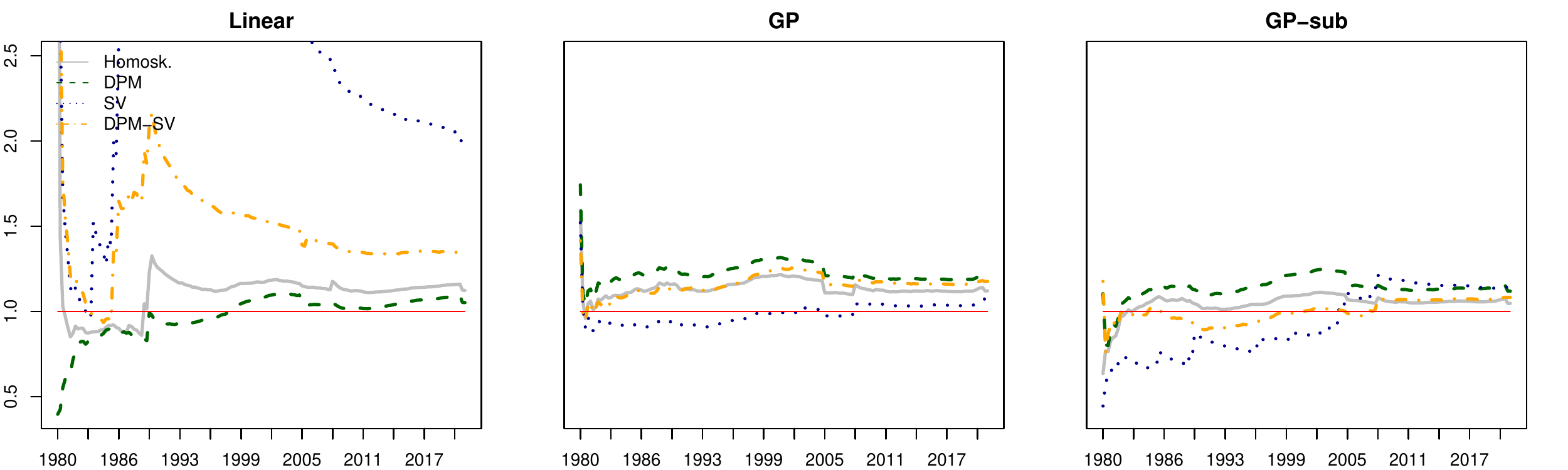}
    \end{minipage}
        \begin{minipage}[c]{1\linewidth}
        \vspace{0.2cm}
       \textbf{(a) 4-quarters-ahead}
    \end{minipage}
    \begin{minipage}[c]{1\linewidth}
           \includegraphics[scale=0.51]{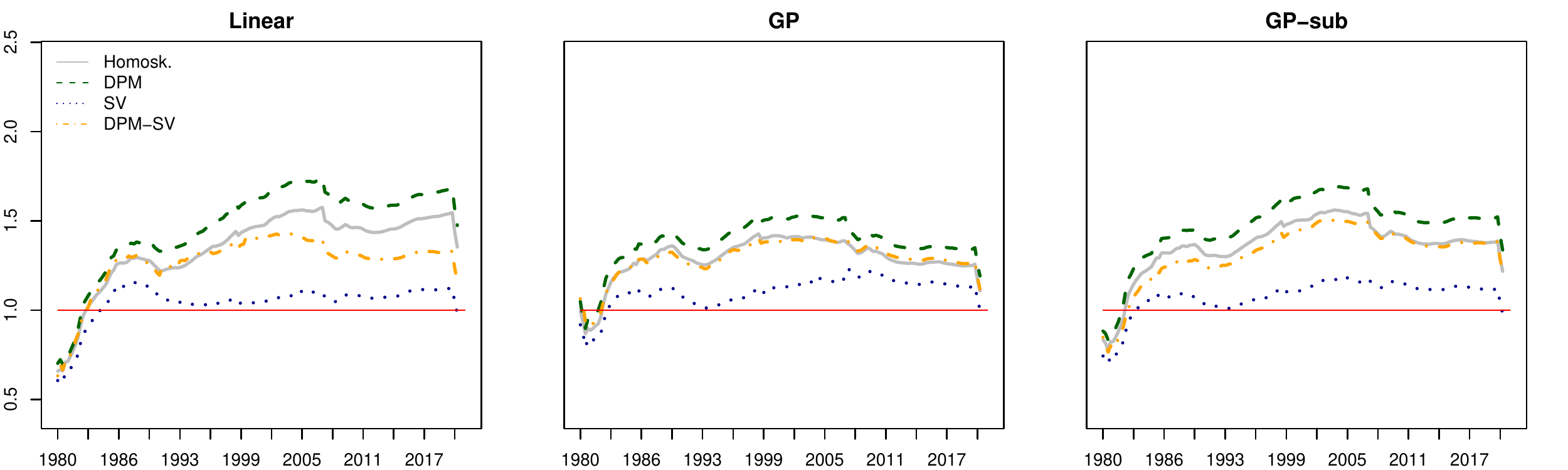}
    \end{minipage}
    \caption{Cumulative QS(0.95) against the UC-SV model: Moderately sized models}
    \label{fig:cumQS_small}
\end{figure}
\begin{figure}[h]
    \centering
    \begin{minipage}[c]{1\linewidth}
       \textbf{(a) 1-quarter-ahead}
    \end{minipage}
    \begin{minipage}[c]{1\linewidth}
           \includegraphics[scale=0.51]{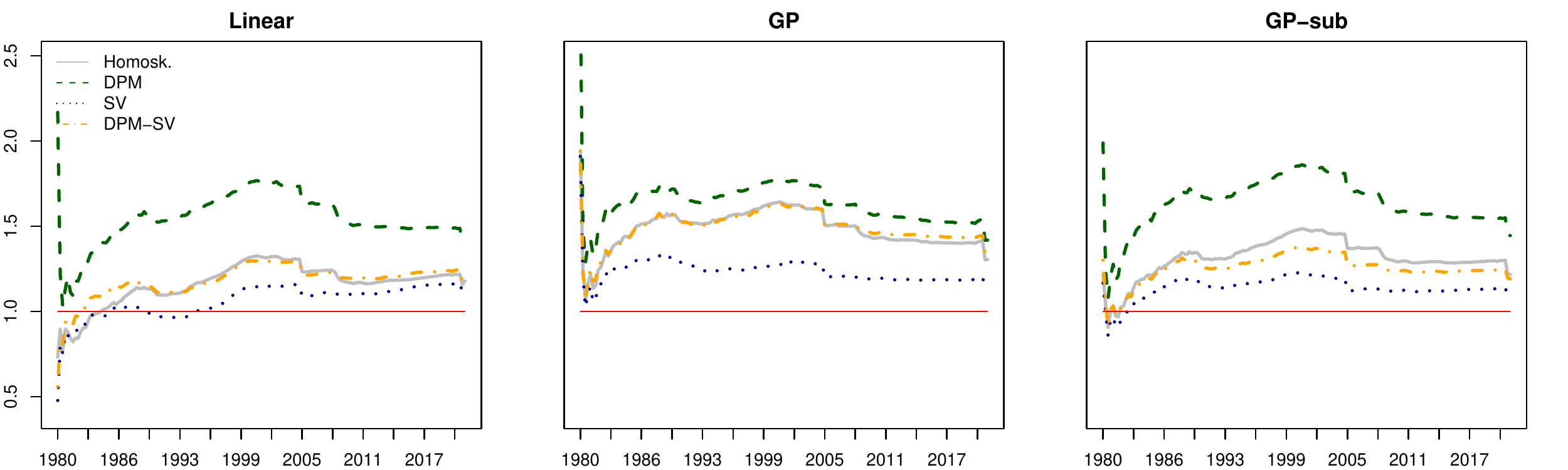}
    \end{minipage}
        \begin{minipage}[c]{1\linewidth}
        \vspace{0.2cm}
       \textbf{(a) 4-quarters-ahead}
    \end{minipage}
    \begin{minipage}[c]{1\linewidth}
           \includegraphics[scale=0.51]{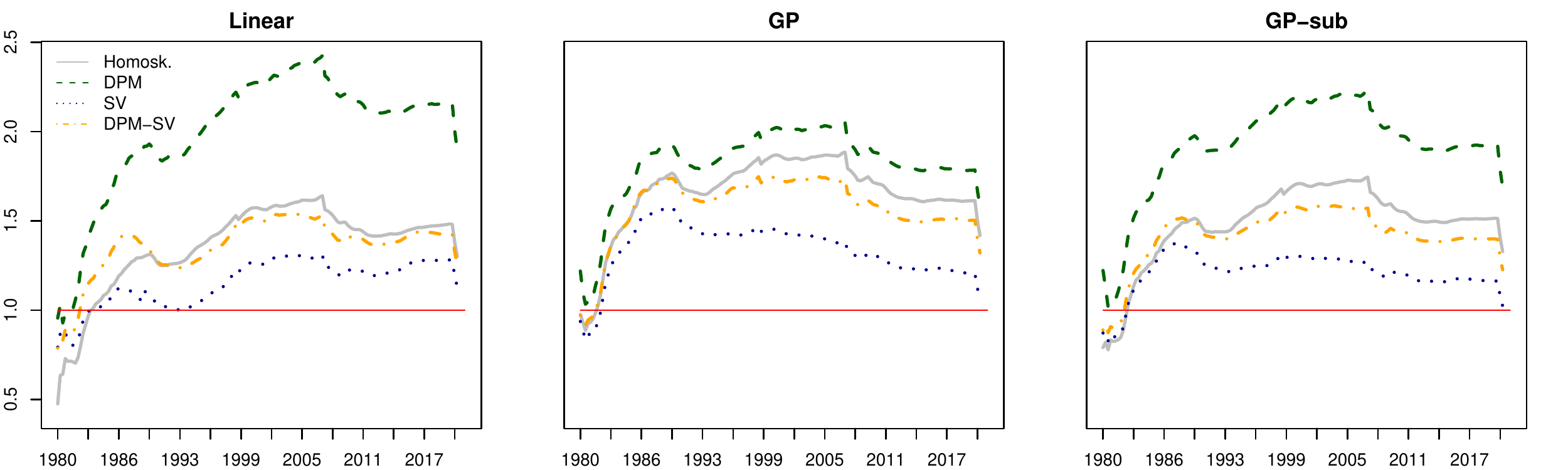}
    \end{minipage}
    \caption{Cumulative QS (0.95) against the UC-SV model: Large models}
    \label{fig:cumQS_large}
\end{figure}

\end{appendices}

\end{document}